\documentclass[apj]{emulateapj}
\usepackage{graphicx}
\usepackage{subfigure}
\usepackage[usenames]{color}
\usepackage{longtable}

\newcommand{\boldsymbol}[1]{\mbox{\boldmath{${#1}$}}}

\newcommand{\bd}{\begin{displaymath}}
\newcommand{\ed}{\end{displaymath}}
\newcommand{\be}{\begin{equation}}
\newcommand{\ee}{\end{equation}}
\newcommand{\beaa}{\begin{eqnarray*}}
\newcommand{\eeaa}{\end{eqnarray*}}
\newcommand{\bea}{\begin{eqnarray}}
\newcommand{\eea}{\end{eqnarray}}
\newcommand{\Ha}{H$\alpha$}
\newcommand{\Hb}{H$\beta$}
\def \Lya{\ensuremath{\mathrm{Ly}\alpha\ }}

\newcommand\HST{\textit{HST}}

\shorttitle{The story of supernova `Refsdal' told by MUSE}
\shortauthors{Grillo et al.}

\begin{document}


\title{The story of supernova `Refsdal' told by MUSE$^\star$}


\author{C.~Grillo\altaffilmark{1}, W.~Karman\altaffilmark{2}, S.~H.~Suyu\altaffilmark{3}, P.~Rosati\altaffilmark{4}, I.~Balestra\altaffilmark{5}, A.~Mercurio\altaffilmark{6}, M.~Lombardi\altaffilmark{7}, T.~Treu\altaffilmark{8}, G.~B.~Caminha\altaffilmark{4}, A.~Halkola, S.~A. Rodney\altaffilmark{9,10,11}, R.~Gavazzi\altaffilmark{12}, K.~I.~Caputi\altaffilmark{2}}

\email{grillo@dark-cosmology.dk}


\altaffiltext{$\star$}{ This work is based in large part on data
  collected at ESO VLT (prog.ID 294.A-5032) and NASA \HST.}
\altaffiltext{1}{Dark Cosmology Centre, Niels Bohr Institute,
  University of Copenhagen, Juliane Maries Vej 30, DK-2100 Copenhagen,
  Denmark} 
\altaffiltext{2}{Kapteyn Astronomical Institute, University of Groningen, Postbus 800, 9700 AV Groningen, the Netherlands}
\altaffiltext{3}{Institute of Astronomy and Astrophysics, Academia
  Sinica, P.O. Box 23-141, Taipei 10617, Taiwan}
\altaffiltext{4}{Dipartimento di Fisica e Scienze della Terra,
  Universit\`a degli Studi di Ferrara, Via Saragat 1, I-44122 Ferrara,
  Italy} 
\altaffiltext{5}{University Observatory Munich, Scheinerstrasse 1, D-81679 Munich, Germany}
\altaffiltext{6}{INAF - Osservatorio Astronomico di Capodimonte, Via
  Moiariello 16, I-80131 Napoli, Italy} 
\altaffiltext{7}{Dipartimento di Fisica, Universit\`a  degli Studi di
  Milano, via Celoria 16, I-20133 Milano, Italy}
\altaffiltext{8}{Department of Physics and Astronomy, University of California, Los Angeles, CA 90095}
\altaffiltext{9}{Department of Physics and Astronomy, University of South Carolina, 712 Main St., Columbia, SC 29208, USA}
\altaffiltext{10}{Department of Physics and Astronomy, The Johns Hopkins University, 3400 N. Charles St., Baltimore, MD 21218, USA}
\altaffiltext{11}{Hubble Fellow}
\altaffiltext{12}{Institut d'Astrophysique de Paris, UMR7095 CNRS-Universit\`e Pierre et Marie Curie, 98bis bd Arago, F-75014 Paris, France}

\begin{abstract}
We present Multi Unit Spectroscopic Explorer (MUSE) observations in the core of the Hubble Frontier Fields (HFF) galaxy cluster MACS J1149.5+2223, where the first magnified and spatially-resolved multiple images of supernova (SN) `Refsdal' at redshift 1.489 were detected. Thanks to a Director's Discretionary Time program with the Very Large Telescope and the extraordinary efficiency of MUSE, we measure 117 secure redshifts with just 4.8 hours of total integration time on a single 1 arcmin$^{2}$ target pointing. We spectroscopically confirm 68 galaxy cluster members, with redshift values ranging from 0.5272 to 0.5660, and 18 multiple images belonging to 7 background, lensed sources distributed in redshifts between 1.240 and 3.703. 
Starting from the combination of our catalog with those obtained from extensive spectroscopic and photometric campaigns using the \textit{Hubble Space Telescope}, we select a sample of 300 (164 spectroscopic and 136 photometric) cluster members, within approximately 500~kpc from the brightest cluster galaxy, and a set of 88 reliable multiple images associated to 10 different background source galaxies and 18 distinct knots in the spiral galaxy hosting SN `Refsdal'. We exploit this valuable information to build 6 detailed strong lensing models, the best of which reproduces the observed positions of the multiple images with a root-mean-square offset of only 0.26\arcsec. We use these models to quantify the statistical and systematic errors on the predicted values of magnification and time delay of the next emerging image of SN `Refsdal'. We find that its peak luminosity should occur between March and June 2016, and should be approximately 20\% fainter than the dimmest (S4) of the previously detected images but above the detection limit of the planned \HST/WFC3 follow-up. We present our two-dimensional reconstruction of the cluster mass density distribution and of the SN `Refsdal' host galaxy surface brightness distribution. We outline the roadmap towards even better strong lensing models with a synergetic MUSE and \HST\ effort.

\end{abstract}





\section{Introduction}
\label{sec:intro}

Massive clusters of galaxies have become a key place for investigating
the formation of large scale structures because of their rich dark matter content (e.g., \citealt{new13}). They also serve as a unique laboratory for the study of galaxy formation and evolution (e.g., \citealt{ann14,ann15}). Owing to the ability of massive clusters to act as powerful gravitational lenses, observations of galaxy clusters simultaneously yield important clues about their mass distribution and extend the limits of detectability of faint background sources (e.g., \citealt{bal13}; \citealt{mon14}). In this context, large observational programs have recently been carried out or are underway from space (e.g., the Cluster Lensing And Supernova survey with Hubble, CLASH \citealt{pos12}, and Hubble Frontier Fields, HFF PI: J.~Lotz) and from the ground with wide-field imaging and dedicated multi-object spectroscopy (e.g., the CLASH-VLT Large Programme, \citealt{ros14}, and several Keck programs). Similar efforts are underway at X-ray wavelengths to provide complementary information on the hot gas component (e.g., \citealt{ogr15}).

The Grism Lens-Amplified Survey from Space (GLASS, GO-13459; PI: T.~Treu) obtained deep slitless near-IR spectra with \HST\ for ten galaxy clusters imaged by the HFF or CLASH programs (\citealt{tre15}). MACS J1149.5+2223 (hereafter MACS 1149), located at $z = 0.542$, is one of them and it has been the target of several strong lensing studies (\citealt{smi09}; \citealt{zit09,zit11}; \citealt{zhe12}; \citealt{bra14}; \citealt{rau14}; \citealt{RichardEtal14}; \citealt{jon14}; \citealt{coe15}; \citealt{sha15}; \citealt{ogu15}; \citealt{die15}; \citealt{jau15}; \citealt{kaw15}). In November 2014, \HST\ imaging of the MACS~1149 cluster from the GLASS program revealed four images of a strongly lensed supernova (SN) in a multiply-imaged and highly-magnified spiral galaxy at $z = 1.489$ (\citealt{kel15}). This object, nicknamed SN `Refsdal', is the first SN to be resolved into four multiple images, forming an almost perfect Einstein cross around an elliptical cluster member of MACS 1149. The SN `Refsdal' host is lensed into two additional complete multiple images by the galaxy cluster, and a fifth appearance of SN `Refsdal' has recently been observed in one of those host galaxy images (\citealt{kel16}). The final host galaxy image is a leading image, and SN `Refsdal' likely appeared there more than 10 years earlier, unobserved. After the discovery of `Refsdal', a number of follow-up programs have been triggered, aimed at classifying the SN as well as measuring the time delays. These include deep spectroscopic follow-up observations with the \HST\ WFC3-IR grism (GO-14041; PI: P.~Kelly), which is considerably deeper in the G141 grism than the GLASS data taken at discovery. Also, a multicolor photometric light curve has been obtained using follow-up \HST\ imaging as part of the HFF imaging campaign and target of opportunity follow-up from the FrontierSN program (GO-13790; PI: S.~Rodney). An ongoing \HST\ imaging campaign will continue to monitor the field in the coming year (GO-14199; PI: P.~Kelly). Precise measurements of time delays and magnifications between the multiple images of the SN can provide valuable information either about the lens itself or about the path of the lensed light through the Universe.  Although the programs to observe the SN `Refsdal' images are still ongoing, current observations have already been used to test the accuracy of strong lensing models for MACS 1149 (see \citealt{tre15b}; \citealt{kel16}; \citealt{rod16}).  Alternatively, if one adopts a given mass model that has sufficient precision, the SN `Refsdal' time delays can be used to constrain the expansion history of the Universe, without intermediate calibrations  (e.g., \citealt{ref64}; \citealt{suy14}).

Measuring the inner dark matter density profile of clusters is among the best methods to test $\Lambda$CDM predictions and characterize the complex interplay between baryons in the central cD galaxy, its supermassive black hole, and the intracluster medium (hot gas and stars). The physics entering present-day hydrodynamical numerical simulations is starting to capture these effects (e.g., \citealt{dub13}; \citealt{lap15}; \citealt{mar14}) and predictions regarding halo contraction/expansion, dissipative (in situ) star formation, and non-collisional accretion processes can be directly tested with mass models derived from spatially resolved stellar kinematics of the brightest cluster galaxies (BCGs) (e.g., \citealt{mir95}; \citealt{san04}; \citealt{new13a}). In addition, the dynamics of cluster members allows us to reduce uncertainties in lensing mass such as those due to the assumption of the cluster members following the Faber-Jackson relation.  Since the relation has a $\sim30\%$ intrinsic scatter between light and total mass, better proxies for the total mass of cluster members (i.e., stellar kinematics) would alleviate a primary source of systematic errors in magnification maps and time-delay predictions (e.g., \citealt{dal11}; \citealt{mon15}). The large spectroscopic data set available for MACS~1149 will make it the ideal laboratory to address all the previously mentioned topics, as partly anticipated by this work.

In January 2015, we proposed observations at the Very Large Telescope (VLT) with the Multi Unit Spectroscopic Explorer (MUSE) in MACS 1149, while the four multiple images of the SN `Refsdal' were still glowing (as confirmed by the \HST\ monitoring campaign). The combination of WFC3-IR-GRISM and MUSE observations was intended to provide a highly complementary data set: integral field spectroscopy with complete wavelength coverage from 0.47-1.7$\,\mu$m, with both near-IR high {\em spatial} resolution (from \HST/WFC3-IR-GRISM) and optical high {\em spectral} resolution (from VLT/MUSE) over the entire cluster core. The observations were time critical because an accurate mass model of the lens galaxy cluster was needed to predict robustly the time delays of all the multiple images of SN `Refsdal' and plan future observations to see the rise of its next occurrence and obtain a precise time-delay measurement of this unique lensing event. The exceptional suitability and power of MUSE to address these scientific objectives had been previously demonstrated in our similar Science Verification program 60.A-9345 (\citealt{kar15}) on Abell S1063, another HFF massive lens galaxy cluster with multi-color \HST\ data, and in the commissioning program 60.A-9100 on SMACS J2031.8$-$4036 (\citealt{ric15}).

Five different groups, including ours, have worked together to merge all available follow-up data of MACS~1149 and decide which information to use in their independent strong lensing analyses. A description of this collaboration and a comparison between model forecasts for SN `Refsdal' are given by \citet{tre15b}. Our contribution, in terms of both spectroscopic data set and strong lensing modeling, is only briefly summarized in that work and is illustrated more comprehensively here.

This paper is organized as follows. In Section 2, we introduce the \HST\ data used in this work. In Section 3, we present our MUSE spectroscopic observations of MACS~1149 and provide a complete redshift catalog. We describe our strong lens modeling of the cluster in Section 4, focusing in particular on the model-predicted quantities for the multiple images of SN `Refsdal'. We draw our conclusions in Section 5.

Throughout this paper, we assume a $\Lambda$CDM cosmology with
$H_{0}=70$ km s$^{-1}$ Mpc$^{-1}$, $\Omega_{\rm m}=0.3$, and $\Omega_{\Lambda}=0.7$.  
In this cosmology, $1''$ corresponds to 6.36\,kpc at the MACS 1149
redshift of $z=0.542$. All magnitudes are given in the AB
system. Parameter estimates are provided as the median values with
statistical uncertainties given by the 16th and 84th percentiles (corresponding to
68\% confidence levels (CLs)) unless otherwise stated.

\section{\HST\ imaging}

As part of the CLASH sample, MACS~1149 was observed in \HST\
Cycle 18, between December 2010 and March 2011, in 15 broadband
filters from 0.2 to 1.7 $\mu$m, to a total depth of 18 orbits, to which previous archival \HST\ images in the F814W band were added (see \citealt{pos12}; \citealt{zhe12}; \citealt{jou14}). The images were processed with standard calibration techniques and combined using drizzle algorithms to generate mosaics with pixel scales of 0.030\arcsec$\,$ and 0.065\arcsec$\,$ (see \citealt{koe07,koe11} for all the details). The values of exposure times, limiting magnitudes, and extinction coefficients in each filter are provided by \citet{pos12} and \citet{bra14}. In the following analysis (except for Figure \ref{fi02}), we will use the \HST\ mosaics with 0.065\arcsec$\,$ pixel scale in the 16 CLASH bands.

\section{VLT spectroscopy}
\label{sec:spec}

The MUSE (\citealt{bac12}) 
instrument, mounted on the VLT at the Paranal observatory,
is the ideal instrument to spectroscopically 
study the galaxies residing in the central regions of galaxy clusters and, simultaneously, those in the foreground and background (for example, see \citealt{kar15}). The power of MUSE stems from its field of view (FoV) of 1 arcmin$^{2}$, its high spatial resolution (0.2\arcsec), 
large spectral coverage (4750-9350 \AA), and relatively high spectral 
resolution ($R\sim3000$). We targeted MACS 1149 with MUSE for 6 hours, using Director's Discretionary Time (DDT) in service 
mode (prog.ID 294.A-5032), centered at $\alpha$~=~11:49:35.75, $\delta$~=~+22:23:52.4 (PI: C.~Grillo).

MACS 1149 was observed with MUSE on the 14th of 
February 2015 for one hour, on the 
21st of March 2015 for four hours, and on the 
12th of April 2015 for one hour. Each hour consisted of two exposures of 1440s, so that 
the total exposure time adds up to 17280s or 4.8 hours. The observational conditions 
were clear and photometric, with a seeing of less than $1.1\arcsec$ in 10
out of the 12 exposures. The two
exposures in March were executed 
under a significantly higher seeing of $\sim2\arcsec$. 

We applied a dithering pattern between different exposures,
where each exposure was offset by a fraction of an arcsecond in right 
ascension and declination. The observations started with a position angle of 4$^{\circ}$,
and we rotated every subsequent exposure by 90$^{\circ}$ (the approximate FoV of the MUSE observations in MACS~1149 is shown in Figure 2 of \citealt{tre15b}). In this way, we have 3 
exposures at each rotation angle. Using a bright star 
in the FoV, we measure a resulting full width half maximum (FWHM) of $0.9\arcsec$ in the 
final datacube, which is only slightly larger (difference $<0.15\arcsec$) than the FWHM  
we obtain when we only include the observations where our observational conditions were fulfilled.

\subsection{Data reduction}
The data reduction was performed using the MUSE Data Reduction Software version
1.0\footnote{http://www.eso.org/sci/software/pipelines/muse/}. This pipeline subtracted the bias from every exposure, applied a 
flatfielding, and calibrated the wavelength using 70 different arc lamp lines. 
Further, we used the illumination files to correct for the illumination patterns
in the data, and the line spread function and astrometry of the instrument are 
taken from recent calibration files. All exposures are flux calibrated by the 
pipeline using observations of a standard star observed in the same night as 
the exposure. The pipeline then models the sky, and
subtracts a model of the sky from every exposure individually. As small offsets
 of the different exposures will reduce the quality of the final datacube, we 
align each exposure by measuring the positions of multiple bright objects in the 
{\em HST} images and the exposure, and apply an offset that aligns the images.
As demonstrated by \citet{kar15}, this results in a positional accuracy 
less than one pixel. The reduced pixel tables are then combined and resampled 
to a datacube with a grid of 0.2$\arcsec$ in the spatial directions, and 
1.25 \AA\ per wavelength step. In addition, we create a more finely sampled datacube with spectral pixels of 0.82 \AA. The original sampling datacube is used to robustly determine emission lines in low S/N sources, while the finely-sampled datacube is employed to fit emission lines with higher precision. An additional advantage  of a higher sampling is that the wavelength range between skylines is sometimes better resolved, and allows for a clearer detection of emission lines.

After the standard data reduction by the pipeline, we remove the remainder of 
sky signal with a {\sc python} script. This script fits a Gaussian to the 
distribution of pixels in 11 blank regions in the sky, and subtracts 
the central value of the Gaussian distribution 
from the complete observed field. We repeat this process for every 
wavelength bin of our datacube, and mask out any wavelength where the dispersion
 of the sky, as measured by the width of the Gaussian, is larger than 
$3 \times 10^{-2}$ erg s$^{-1}$ cm$^{-2}$ \AA$^{-1}$. This procedure 
effectively removes any sky residual, while masking out wavelengths where sky 
lines dominate the signal.

To determine the position of sources with a bright enough continuum, we used 
the pipeline to stack the datacube in the spectral direction, after creating
a spectral mask for sky lines. We then ran {\sc sextractor} (\citealt{ber96})
on this image to obtain the positions of these sources. As a next step, we 
extracted a one dimensional spectrum at each position, by summing all spatial 
pixels within a 0.6$\arcsec$ radius, resulting in 92 spectra. We added 
21 more sources that were visible in the stacked image, but which were
missed by {\sc sextractor} due to blending or faintness.

Although most galaxies with strong emission lines are also detected with this
method, several are not bright enough. In order to look for these galaxies,
we performed a visual inspection of the datacube, and found 21 additional
sources showing only emission lines. Therefore, the total number of spectra
is 134.

\subsection{Spectral analysis}

We inspected each extracted spectrum, and determined redshifts, $z_{\rm sp}$, for 129 out of 
134 objects (see Table \ref{redshifts}). Given the spectral resolution of MUSE and the number of identified lines in the spectra, we decided to provide redshift values with a precision in the fourth decimal place up to the cluster members (i.e., $z_{\rm sp} \lesssim 0.6$). To indicate the reliability of a redshift measurement, we assigned to each redshift a quality flag (QF) as follows: 4 or extremely secure ($\delta z_{\rm sp}<$~0.0004), 3 or very secure ($\delta z_{\rm sp}<$~0.001), 2 or secure ($\delta z_{\rm sp}<$~0.01), 1 or uncertain, 9 or secure from a single emission line. These error estimates include systematics uncertainties due to different methods of redshift measurements, as estimated using independent measurements by two of us. One method is based on a first cross-correlation with a template using {\sc SpecPro} (\citealt{mas11}), with the use of an additional GMASS template at higher resolution (\citealt{kur13}), and following visual check and manual fine-tuning to center possible emission and absorption features, in case of residual offsets. The second method does not rely on any templates, but purely on fitting a set of lines visually detected in a spectrum (when possible, the fit of a line is made with a gaussian to determine more robustly its center). Redshift measurements with QF greater than 1 are considered secure. According to this criterion, we have 117 objects with secure redshifts\footnote{Due to a subtlety in our quality flags, three sources (i.e., 93, 98, and 112c) are classified differently here from that in \citet{tre15b}.}. In Table \ref{redshifts}, we list previous redshift estimates, $z_{\rm sp,pre.}$, for some of the objects in our catalog. In detail, we compare our results with those of \citet{ebe14} (labeled E) and \citet{smi09} (labeled S). We also record the presence (Y) of emission lines (EL) in the spectra.

We note that while writing this paper, the MUSE data and
redshifts of MACS~1149 were also presented by \citet{jau15}, based on a 4 hour fraction of the total MUSE exposure. Although there is a significant overlap between the redshifts presented by us and the other group,
we provide here a larger and more complete catalog which includes redshift measurement uncertainties. In detail, we report the redshift values of 77 cluster members, 7 foreground galaxies, 25 singly-imaged and 7 multiply-imaged background galaxies versus, respectively, 55, 6, 17 and 4 objects in \citet{jau15}. There is fairly good agreement between the redshift values of the two samples, but we find three objects with incompatible redshifts. Based on our analysis (see below), these are a star, a foreground galaxy and a cluster member (objects 2, 8 and 50 in Table \ref{redshifts}), of which the first two are listed as foreground galaxies, at significantly different redshifts, and the last one as a cluster member, at a slightly lower $z$, by \citet{jau15} (objects 75, 76 and 3, respectively, in that work). In total, we provide a complete redshift list of 118 objects with QF, exploiting our full set of MUSE observations.

\begin{figure}
\centering
\includegraphics[width=0.48\textwidth]{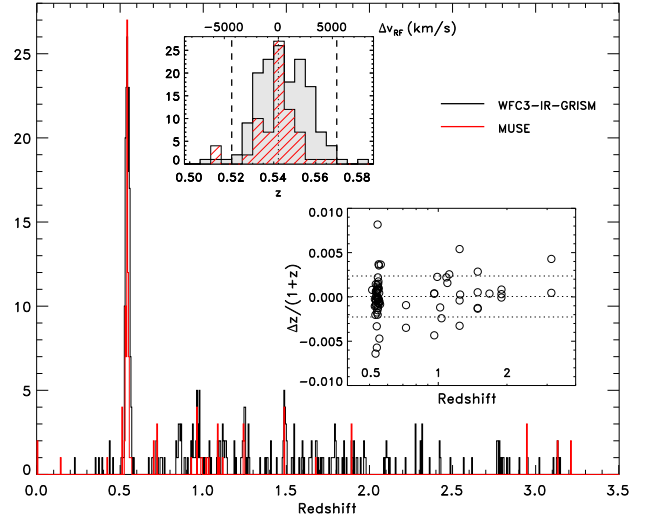}
\caption{Spectroscopic redshift distribution of the objects detected in the
core of MACS 1149 from the MUSE ($N=117$) and WFC3-IR-GRISM ($N=389$) catalogs. The top inset
shows a zoom-in around the cluster redshift $z$ = 0.542 (red-hatched
histogram from MUSE, shaded from WFC3-IR-GRISM); top axis gives the
relative 
rest-frame velocity, vertical lines are the redshifts [0.520,
0.570] which include the 68 MUSE cluster members found from a kinematical
analysis. MUSE and WFC3-IR-GRISM redshifts are compared in the bottom inset,
showing also mean and standard deviation of the distribution 
(horizontal dashed lines).
}
\label{fi04}
\end{figure}

In Figure \ref{fi04}, we show the redshift distribution of the MUSE and WFC3-IR-GRISM secure redshifts. 
There are 74 redshifts in common between MUSE and WFC3-IR-GRISM, the differences are remarkably small, with a standard deviation of 0.0023 in $(z_{\rm WFC3-IR-GRISM}-z_{\rm MUSE})/(1+z_{\rm MUSE})$ (see inset in Figure \ref{fi04}), consistent with the uncertainties expected given the spectral resolution of the WFC3-IR-GRISM data. 

In this section, we illustrate the 
spectral properties of the galaxies, divided into three subsamples: cluster 
members, foreground galaxies and stars, and lensed background galaxies. In 
the last part of this section, we will discuss the multiply lensed images 
into some more detail.


\begin{longtable}{ccccccc}
\caption{MUSE spectroscopic catalog.} \\
\hline \hline \noalign{\smallskip}
 ID & R.A.  & Decl. & $z_{\rm sp}$ & QF & $z_{\rm sp,pre.}$ & EL\\
 & (J2000) & (J2000) & & & & \\
\noalign{\smallskip} \hline \noalign{\smallskip}

1 & 11:49:35.325 & +22:23:37.48 & 0.0000 & 4 & --- & N\\
2 & 11:49:36.973 & +22:23:31.59 & 0.0000 & 4 & --- & N\\
3 & 11:49:38.026 & +22:24:18.64 & 0.1422 & 3 & --- & Y\\
4 & 11:49:35.295 & +22:24:15.97 & 0.4242 & 4 & --- & Y\\
5 & 11:49:34.920 & +22:24:14.57 & 0.5132 & 3 & --- & Y\\
6 & 11:49:34.390 & +22:24:00.96 & 0.5134 & 4 & --- & Y\\
7 & 11:49:34.297 & +22:24:14.45 & 0.5137 & 4 & --- & Y\\
8 & 11:49:34.392 & +22:23:58.67 & 0.5137 & 4 & --- & Y\\
9 & 11:49:37.097 & +22:23:47.15 & 0.5172 & 1 & --- & N\\
\noalign{\smallskip} \hline \noalign{\smallskip} 
10 & 11:49:33.838 & +22:24:06.03 & 0.5272 & 4 & 0.527 (E) & N\\
11 & 11:49:37.510 & +22:24:19.40 & 0.5277 & 3 & --- & N\\
12 & 11:49:37.802 & +22:23:56.80 & 0.5303 & 2 & --- & Y\\
13 & 11:49:36.854 & +22:23:46.97 & 0.5310 & 4 & --- & N\\
14 & 11:49:33.901 & +22:23:33.81 & 0.5310 & 4 & 0.531 (E) & Y\\
15 & 11:49:36.626 & +22:23:46.22 & 0.5325 & 4 & --- & N\\
16 & 11:49:37.306 & +22:23:52.33 & 0.5330 & 4 & 0.533 (E) & N\\
17 & 11:49:36.290 & +22:24:01.26 & 0.5332 & 4 & --- & N\\
18 & 11:49:34.289 & +22:23:49.58 & 0.5332 & 4 & 0.533 (E) & N\\
19 & 11:49:36.736 & +22:24:15.76 & 0.5337 & 4 & 0.534 (E) & N\\
20 & 11:49:33.863 & +22:24:17.65 & 0.5337 & 4 & --- & N\\
21 & 11:49:35.587 & +22:24:11.33 & 0.5337 & 1 & --- & N\\
22 & 11:49:34.043 & +22:24:18.96 & 0.5340 & 4 & --- & N\\
23 & 11:49:35.919 & +22:23:58.55 & 0.5355 & 4 & --- & N\\
24 & 11:49:35.653 & +22:23:23.25 & 0.5357 & 4 & 0.534 (E) & N\\
25 & 11:49:35.095 & +22:23:48.76 & 0.5362 & 1 & --- & N\\
26 & 11:49:34.801 & +22:23:45.54 & 0.5377 & 3 & --- & N\\
27 & 11:49:33.925 & +22:24:03.98 & 0.5382 & 4 & --- & N\\
28 & 11:49:34.517 & +22:24:08.25 & 0.5382 & 3 & --- & N\\
29 & 11:49:35.624 & +22:24:19.36 & 0.5382 & 3 & --- & N\\
30 & 11:49:36.962 & +22:24:07.60 & 0.5389 & 3 & --- & N\\
31 & 11:49:35.684 & +22:23:32.36 & 0.5400 & 4 & --- & N\\
32 & 11:49:35.257 & +22:23:32.54 & 0.5402 & 2 & --- & N\\
33 & 11:49:34.472 & +22:23:41.41 & 0.5402 & 1 & --- & N\\
34 & 11:49:35.152 & +22:23:52.61 & 0.5402 & 1 & --- & N\\
35 & 11:49:37.805 & +22:24:11.01 & 0.5402 & 4 & --- & N\\
36 & 11:49:36.182 & +22:23:46.59 & 0.5407 & 3 & --- & N\\
37 & 11:49:35.696 & +22:23:54.66 & 0.5410 & 4 & --- & N\\
38 & 11:49:36.246 & +22:23:52.39 & 0.5410 & 4 & --- & N\\
39 & 11:49:34.686 & +22:24:02.35 & 0.5412 & 3 & --- & N\\
40 & 11:49:37.241 & +22:23:59.18 & 0.5412 & 3 & --- & N\\
41 & 11:49:37.294 & +22:23:29.87 & 0.5412 & 2 & --- & N\\
42 & 11:49:36.096 & +22:23:53.51 & 0.5412 & 3 & --- & N\\
43 & 11:49:35.007 & +22:23:36.75 & 0.5413 & 4 & 0.541 (E) & N\\
44 & 11:49:36.574 & +22:23:52.72 & 0.5414 & 2 & 0.540 (E) & N\\
45 & 11:49:34.747 & +22:23:34.68 & 0.5417 & 3 & 0.541 (E) & N\\
46 & 11:49:36.045 & +22:23:39.95 & 0.5417 & 4 & --- & Y\\
47 & 11:49:35.731 & +22:24:06.55 & 0.5417 & 4 & --- & N\\
48 & 11:49:34.126 & +22:24:04.61 & 0.5419 & 3 & --- & N\\
49 & 11:49:37.742 & +22:23:29.24 & 0.5422 & 2 & --- & N\\
50 & 11:49:35.662 & +22:23:53.05 & 0.5422 & 3 & --- & N\\
51 & 11:49:35.820 & +22:24:21.41 & 0.5422 & 1 & --- & N\\
52 & 11:49:34.865 & +22:24:03.80 & 0.5426 & 4 & --- & N\\
53 & 11:49:33.641 & +22:24:13.76 & 0.5428 & 2 & --- & N\\
54 & 11:49:36.406 & +22:23:55.55 & 0.5432 & 3 & --- & N\\
55 & 11:49:35.531 & +22:24:16.48 & 0.5432 & 1 & --- & N\\
56 & 11:49:34.856 & +22:23:49.13 & 0.5432 & 1 & --- & N\\
57 & 11:49:34.001 & +22:23:26.30 & 0.5437 & 4 & 0.543 (E) & N\\
58 & 11:49:36.886 & +22:23:31.02 & 0.5437 & 4 & --- & N\\
59 & 11:49:36.187 & +22:23:37.54 & 0.5442 & 2 & --- & N\\
60 & 11:49:36.539 & +22:23:59.08 & 0.5443 & 4 & 0.544 (E) & N\\
61 & 11:49:37.646 & +22:23:44.95 & 0.5450 & 4 & --- & N\\
62 & 11:49:35.470 & +22:23:43.65 & 0.5450 & 4 & 0.544 (E) & N\\
63 & 11:49:33.507 & +22:23:33.65 & 0.5452 & 4 & --- & N\\
64 & 11:49:34.810 & +22:23:23.30 & 0.5452 & 2 & 0.546 (E) & N\\
65 & 11:49:34.250 & +22:23:39.86 & 0.5455 & 4 & --- & N\\
66 & 11:49:37.762 & +22:23:41.33 & 0.5462 & 4 & --- & Y\\
67 & 11:49:36.686 & +22:24:07.24 & 0.5462 & 3 & --- & N\\
68 & 11:49:35.913 & +22:24:08.19 & 0.5462 & 1 & --- & N\\
69 & 11:49:35.399 & +22:23:58.32 & 0.5467 & 4 & --- & N\\
70 & 11:49:35.801 & +22:24:03.14 & 0.5472 & 2 & --- & N\\
71 & 11:49:36.895 & +22:24:16.50 & 0.5477 & 4 & --- & N\\
72 & 11:49:35.511 & +22:24:03.78 & 0.5479 & 4 & 0.547 (E) & N\\
73 & 11:49:36.889 & +22:23:20.79 & 0.5487 & 4 & --- & N\\
74 & 11:49:35.501 & +22:24:14.11 & 0.5492 & 2 & --- & N\\
75 & 11:49:37.604 & +22:23:44.21 & 0.5495 & 4 & 0.549 (E) & N\\
76 & 11:49:37.545 & +22:23:22.51 & 0.5495 & 4 & 0.550 (E) & N\\
77 & 11:49:34.239 & +22:23:33.88 & 0.5505 & 4 & --- & N\\
78 & 11:49:35.424 & +22:24:10.00 & 0.5517 & 3 & --- & N\\
79 & 11:49:35.258 & +22:23:35.05 & 0.5520 & 3 & 0.551 (E) & N\\
80 & 11:49:34.268 & +22:23:53.11 & 0.5525 & 4 & 0.553 (E) & N\\
81 & 11:49:35.370 & +22:24:00.88 & 0.5538 & 3 & --- & N\\
82 & 11:49:34.601 & +22:23:42.06 & 0.5542 & 2 & --- & N\\
83 & 11:49:33.471 & +22:23:38.13 & 0.5561 & 4 & 0.556 (E) & N\\
84 & 11:49:35.953 & +22:23:50.18 & 0.5600 & 4 & 0.560 (E) & N\\
85 & 11:49:36.858 & +22:24:20.16 & 0.5602 & 1 & --- & N\\
86 & 11:49:36.964 & +22:24:11.04 & 0.5660 & 4 & --- & Y\\
\noalign{\smallskip} \hline \noalign{\smallskip}          
87 & 11:49:34.813 & +22:23:50.99 & 0.702 & 4 & --- & Y\\
88 & 11:49:35.255 & +22:23:53.02 & 0.702 & 4 & --- & Y\\
89 & 11:49:34.433 & +22:24:04.81 & 0.722 & 4 & --- & N\\
90 & 11:49:33.607 & +22:23:51.12 & 0.722 & 3 & --- & Y\\
91 & 11:49:36.429 & +22:23:36.59 & 0.723 & 3 & --- & Y\\
92 & 11:49:36.904 & +22:23:32.23 & 0.929 & 4 & --- & Y\\
93 & 11:49:33.601 & +22:23:22.23 & 0.959 & 9 & --- & Y\\
94 & 11:49:34.850 & +22:23:24.07 & 0.960 & 4 & --- & Y\\
95 & 11:49:34.711 & +22:23:26.21 & 0.960 & 9 & --- & Y\\
96 & 11:49:34.691 & +22:23:29.75 & 0.960 & 9 & --- & Y\\
97 & 11:49:35.913 & +22:23:32.52 & 0.990 & 9 & --- & Y\\
98 & 11:49:35.402 & +22:23:22.37 & 0.990 & 9 & --- & Y\\
99 & 11:49:35.673 & +22:23:22.01 & 1.019 & 9 & --- & Y\\
100 & 11:49:38.107 & +22:24:11.31 & 1.033 & 4 & --- & Y\\
101 & 11:49:36.242 & +22:24:16.90 & 1.086 & 4 & --- & Y\\
102 & 11:49:35.647 & +22:24:20.62 & 1.088 & 9 & --- & Y\\
103 & 11:49:35.564 & +22:24:19.21 & 1.088 & 9 & --- & Y\\
104 & 11:49:37.198 & +22:24:04.65 & 1.096 & 9 & --- & Y\\
105 & 11:49:37.991 & +22:23:55.73 & 1.117 & 9 & --- & Y\\
106a & 11:49:36.012 & +22:23:38.06 & 1.240 & 9 & --- & Y\\
106b & 11:49:36.679 & +22:23:47.94 & 1.240 & 9 & --- & Y\\
106c & 11:49:36.893 & +22:23:52.08 & 1.240 & 9 & --- & Y\\
107 & 11:49:34.045 & +22:24:00.38 & 1.247 & 4 & --- & Y\\
108 & 11:49:33.770 & +22:23:37.01 & 1.249 & 9 & --- & Y\\
109a & 11:49:35.297 & +22:23:45.99 & 1.489 & 4 & 1.491 (S) & Y\\
109b & 11:49:36.829 & +22:24:08.67 & 1.489 & 4 & 1.491 (S) & Y\\
109c & 11:49:35.900 & +22:23:50.13 & 1.489 & 4 & 1.491 (S) & Y\\
110 & 11:49:34.220 & +22:23:41.74 & 1.489 & 9 & --- & Y\\
111 & 11:49:35.335 & +22:24:22.21 & 1.676 & 9 & --- & Y\\
112a & 11:49:37.455 & +22:23:32.86 & 1.892 & 3 & 1.894 (S) & Y\\
112b & 11:49:37.570 & +22:23:34.55 & 1.892 & 3 & 1.894 (S) & Y\\
112c & 11:49:36.580 & +22:23:23.10 & 1.892 & 9 & 1.894 & Y\\
113a & 11:49:34.322 & +22:23:48.48 & 2.949 & 4 & --- & Y\\
113b & 11:49:34.656 & +22:24:02.74 & 2.949 & 4 & --- & Y\\
113c & 11:49:37.006 & +22:24:22.05 & 2.949 & 4 & --- & Y\\
114 & 11:49:35.635 & +22:23:39.74 & 2.990 & 1 & --- & N\\
115 & 11:49:36.068 & +22:23:43.21 & 3.040 & 1 & --- & N\\
116a & 11:49:34.303 & +22:24:12.02 & 3.130 & 4 & 2.497$^{\star}$ (S) & Y\\
116b & 11:49:33.775 & +22:23:59.40 & 3.130 & 4 & 2.497$^{\star}$ (S) & Y\\
117a & 11:49:36.031 & +22:23:24.67 & 3.216 & 9 & --- & Y\\
117b & 11:49:36.965 & +22:23:34.43 & 3.216 & 9 & --- & Y\\
118a & 11:49:34.000 & +22:24:12.66 & 3.703 & 9 & --- & Y\\
118b & 11:49:33.802 & +22:24:09.53 & 3.703 & 9 & --- & Y\\
\noalign{\smallskip} \hline \noalign{\smallskip}          
119 & 11:49:35.494 & +22:23:49.46 & 999 & 0 & --- & N\\
120 & 11:49:37.470 & +22:23:56.82 & 999 & 0 & --- & N\\
121 & 11:49:34.220 & +22:23:21.91 & 999 & 0 & --- & N\\
122 & 11:49:33.716 & +22:23:37.21 & 999 & 0 & --- & N\\
123 & 11:49:37.059 & +22:23:22.23 & 999 & 0 & --- & N\\
\noalign{\smallskip} \hline \\
\noalign{Quality Flags (QF) in column 5 are: 4 extremely secure ($\delta z <$ 0.0004), 3 very secure ($\delta z <$ 0.001), 2 secure ($\delta z <$ 0.01), 1 uncertain, 9 secure (based on a single emission line). The references for redshifts in column 6 are E for \citet{ebe14} and S for \citet{smi09}. The last column presents the presence of emission lines in the spectra. \newline
$^{\star}$More details about this redshift discrepancy are given in Sect.~3.2.4.}
\label{redshifts}
\end{longtable}

\subsubsection{Cluster members}

To select cluster members, we analyzed the velocity distribution of the 117 objects with secure redshifts by applying the 1D adaptive kernel technique (\citealt{pis93}, as implemented by \citealt{fad96}; \citealt{gir96}). This procedure indicates a peak in the velocity distribution at $z = 0.5422 \pm 0.0006$, consisting of 68 secure cluster members, with a dispersion of $1440_{-130}^{+160}$ km s$^{-1}$. This procedure points also to a possible structure at lower redshift, $z \sim 0.5135$, with four members (objects 5, 6, 7 and 8 in Table \ref{redshifts}). By comparing the \HST\ F160W magnitude to the quality flag, we find secure redshifts up to a magnitude limit of 24.6 mag (see also Figure \ref{fi05}). We cross-correlated our sample with the catalog of \citet{ebe14}, and find
20 galaxies with previously determined redshifts, with differences of at most 
$\Delta z <0.002$. 

\begin{figure*}
\centering
 \includegraphics[width=.98\textwidth]{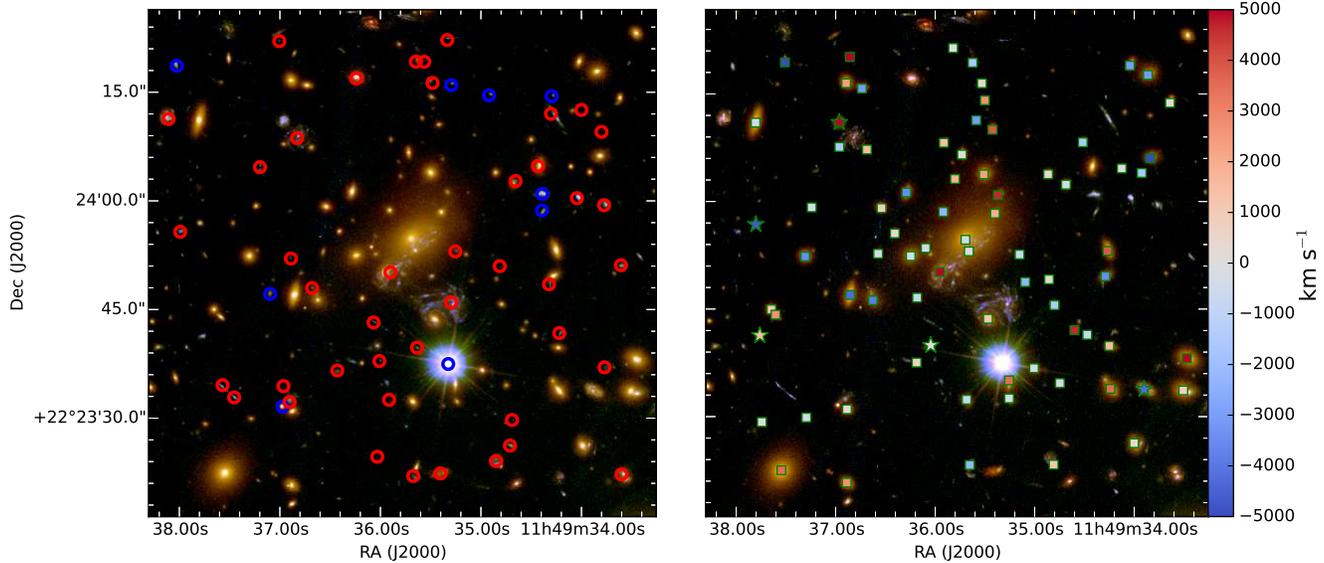}
\caption{Spatial distribution of the objects in the MUSE spectroscopic catalog. 
Foreground ($z_{\rm sp} < 0.5272$) and background ($z_{\rm sp} > 0.5660$) objects are 
identified by blue and red circles, respectively (\emph{on the left}).
Cluster members of MACS 1149 (\emph{on the right}). Points are colored according to 
the values of the relative velocity of each galaxy compared to that of the BCG 
($z_{\rm sp}=0.5410$). Stars represent galaxies with emission lines, while squares 
represent galaxies without any obvious activity (either SF or AGN). \label{fig:cluster}}
\end{figure*}

In Figure \ref{fig:cluster}, we show the distribution of cluster members. There 
is a large number of galaxies North-East of the BCG in the center, and a 
deficit of galaxies in the South-East corner of the field. The relative velocity of the cluster galaxies shows a large spread, though only 2 of the 68 secure members show a rest-frame velocity in excess of 3000 km s$^{-1}$. This is in agreement with previous findings (e.g., \citealt{smi09}) suggesting that the cluster consists of several components and might not be totally relaxed yet. We discuss the structure of the cluster in more detail in the following sections.

Within a projected radius of approximately 200 kpc from the cluster center, we find emission lines in 5 out of 68 member galaxies,  which is a similar fraction
to that found in Abell S1063 (3/34; see \citealt{kar15}). Despite the \Ha\ being outside the
observed wavelength range at this redshift, we detect [\ion{O}{2}], [\ion{O}{3}],
and \Hb\ in two of these five galaxies, while the remaining three only show relatively
weak [\ion{O}{2}] emission. For the three galaxies with only weak [\ion{O}{2}] emission
the source of ionization is likely a small burst of star formation, more diluted onto the continuum emission of an older population (see e.g., \citealt{zab96}; \citealt{pog08}; \citealt{pra09}; \citealt{swi12}). We determine the 
dominating ionization source by comparing the ratio of the [\ion{O}{2}], [\ion{O}{3}],
and \Hb\ lines (\citealt{lam10}). Of the two cluster galaxies 
with all relevant emission lines, one is located 
in the region of the diagram shown by \citet{lam10} where Seyfert 2 and star formation (SF) overlap, while the other is dominated by SF. 

\subsubsection{Foreground objects}

The brightest object in the field is a 16th magnitude star. This star was
used to determine the spatial FWHM of the final product, and to align the
individual exposures. In addition to the bright star, we find a fainter second 
star and seven galaxies in front of the cluster. The second star and six of the
galaxies, appear red in the \HST\ (ACS+WFC3) images, and could be identified as cluster
members if based on their colors only, highlighting the importance of spectroscopic follow up. All but one of the 
foreground galaxies are showing emission lines. The lowest redshift galaxy
is not detected in the continuum and shows \Ha, and weak [\ion{O}{3}] 
$\lambda\lambda$4959,5007 \AA, [\ion{S}{2}], and [\ion{N}{2}] emission, but 
no \Hb\  emission is detected. The second galaxy at $z=0.4242$ shows [\ion{O}{2}],
[\ion{O}{3}], and \Hb\  emission, and using the line ratios from 
\citet{lam10}, we classify it as a galaxy hosting a Seyfert 2 AGN.

The remaining five foreground galaxies are noteworthy. Their redshifts place them
in close proximity, suggesting a physical connection. In addition to their close
distance in the projected velocity space, the four galaxies in the North West have a maximum projected distance between two galaxies of 17.5\arcsec\ 
or 108 kpc. A dynamical analysis (see Sect.~2.2.1) confirms that these galaxies belong to a peak in the velocity distribution at $z \sim 0.5135$. The fifth galaxy, which has an insecure redshift, is located at $\sim$240 kpc
projected distance from the other four. 
For the two galaxies with the smallest separation, the 
{\em HST} images show a disturbed morphology, while the 
small sizes
 of the other two
galaxies prevent us from deriving detailed information on their morphology.
The physical connection of these galaxies is likely related to 
their enhanced state of activity, as groups of interacting 
galaxies are observed and predicted to have increased SF (e.g., \citealt{vai08}; \citealt{mos14}; \citealt{kar15c}; \citealt{emo15}). All four galaxies of this group show a clear set
of emission lines, although the line ratios do not reveal whether SF or nuclear 
activity is dominant. We speculate that we could be observing the last phase of 
activity in this group of galaxies before the dense environment of the group
and later the cluster quenches it.

\subsubsection{Background galaxies}

In addition to the 9 foreground objects and 77 (68 secure) cluster members, we extracted 
43 spectra of galaxies behind the cluster; 
18 out of these 43 belong to seven multiply lensed galaxies, and are discussed below. 


\begin{figure*}
\centering
\includegraphics[width=0.98\textwidth]{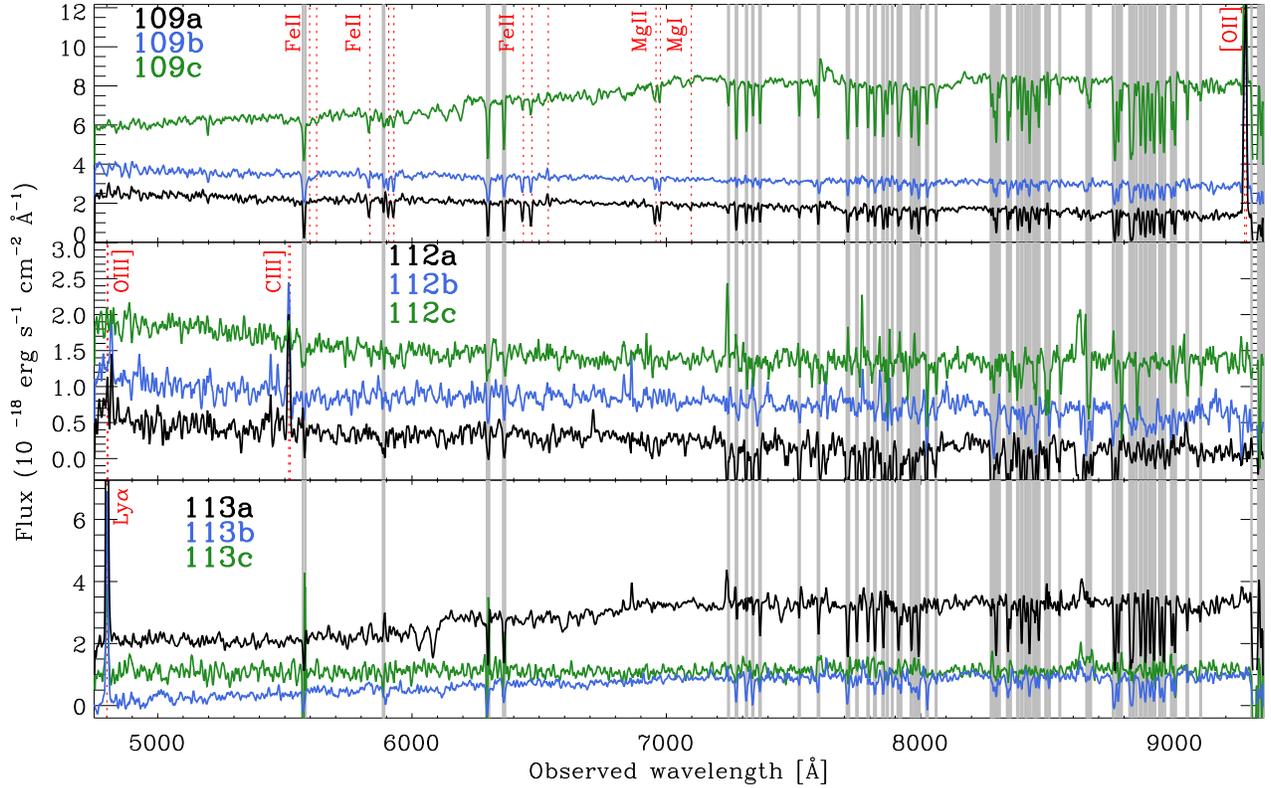}
\caption{Spectra of the multiply lensed galaxies. In each panel, lines of different colors identify the spectra of the multiple images of the same source. Dashed vertical lines locate the redshifted emission or absorption lines detected in the spectra. Gray shaded regions are contaminated by skylines. The 1D spectra were extracted from the MUSE final datacube by summing the flux contribution of all spatial pixels within a circular aperture of 0.6\arcsec\ radius from the object centers. 
}
\label{spec1}
\end{figure*}

\begin{figure*}
\figurenum{\ref{spec1}}
\centering
\includegraphics[width=0.98\textwidth]{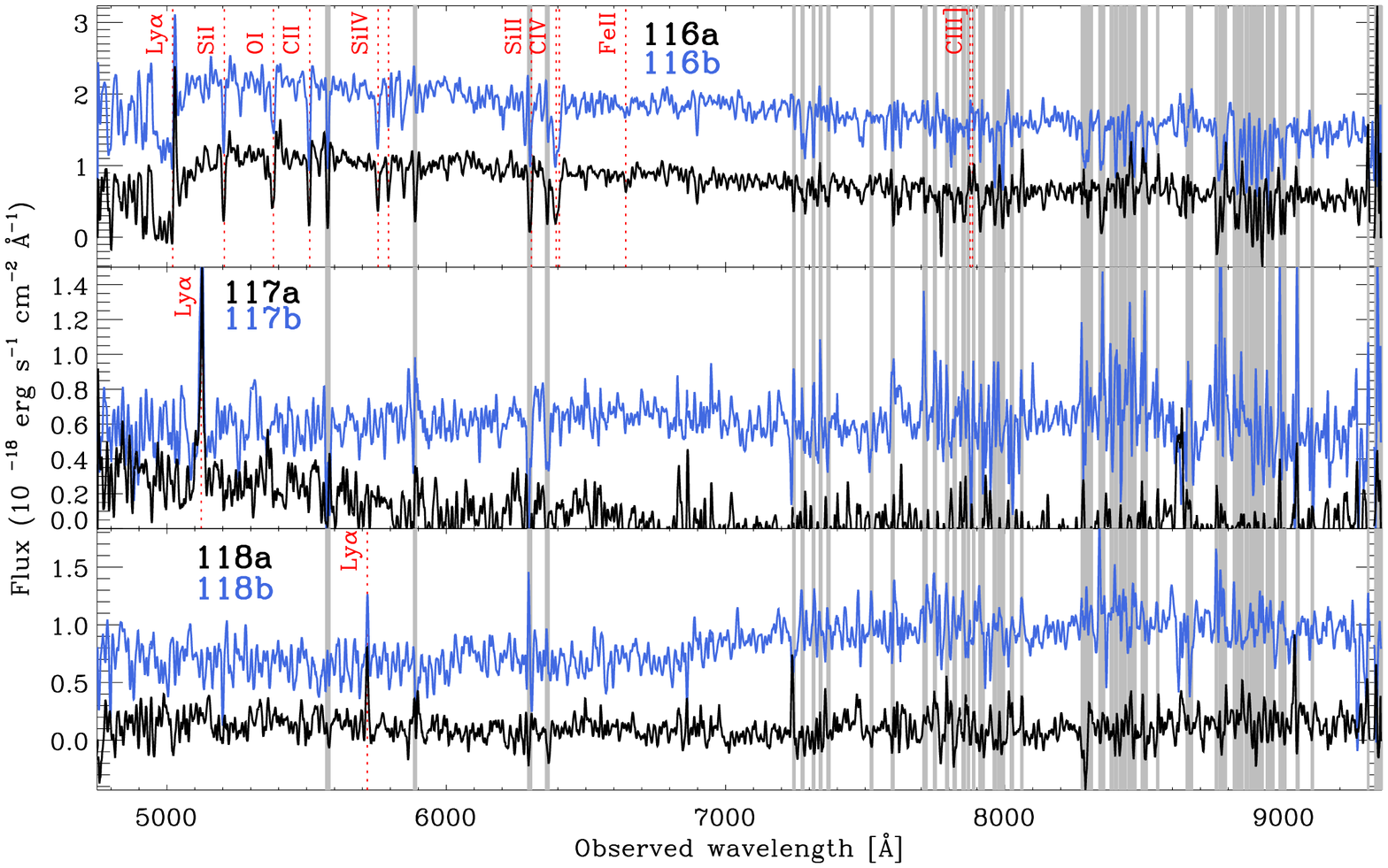}
\caption{\it continued.}
\end{figure*}

\subsubsection{Multiply lensed galaxies}

\citet{jon14} (hereafter J14) identified 12 different galaxies that are very likely
multiply lensed. Our list of extracted spectra contains 18 images, that 
belong to seven of these multiply lensed galaxies. Three of these galaxies
have 
previously determined redshifts 
 (\citealt{smi09}; sources 1, 2, and 3 in J14, 
sources 109, 112, and 116 resp.). Two of these
redshifts agree up to two decimals, and the difference is likely caused
by a difference in resolution. The first galaxy (109) is of particular interest, as this is the host of SN `Refsdal'. We find clear [\ion{O}{2}] 
emission in this galaxy, showing an obvious rotation pattern (see Figures \ref{spec1} and \ref{fi10}). We discuss the 
properties of this galaxy in detail in 
an accompanying paper 
(Karman et al., in prep.). The second galaxy (112) shows two narrow emission doublets, \ion{O}{3}] 
$\lambda\lambda1661,1666$ \AA, and \ion{C}{3}] $\lambda\lambda 1907,1909$ \AA.
The combination of these features sets the redshift securely, and shows
that there is at least a moderate source of ionization present. Unfortunately,
the wavelength range of MUSE prevents us from measuring additional emission 
lines and determining the likely source of ionization.
For the third galaxy (116), we find a significantly
different redshift compared to \citet{smi09}\footnote{This was also noted 
by \citet{jau15}, who posted their work on the arxiv while this paper was being finalized.}. 
While they find $z=2.497$, we measure a redshift
$z=3.130$. We show this spectrum in Figure \ref{spec1}, where a clear
set of UV absorption lines is visible, in addition to \Lya\ in emission. These 
spectral features are detected along the whole arc, and clearly establish the
redshift as $z=3.130$.
The redshifts of these three galaxies were measured independently from HST-WFC3-IR data (\citealt{tre15}; Brammer et al., in prep.) and were found consistent with the MUSE measurements, given the uncertainties.

From our first extraction, we find new redshifts for two multiply lensed
galaxies, which are two \Lya\ emitters (LAEs) at $z=2.949$ and $z=3.216$ (our sources 
113 and 117 resp.). In order to look for weaker emission features,
we extract spectra at all positions of previous photometrically-determined multiply lensed galaxies in our field,
and stack the spectra of images belonging to the same source. We find
one additional clear emission line, using this approach, which is a
LAE at $z=3.703$ (source 118). After determining this redshift,
we found a weak emission line at this wavelength in both of the original
unstacked spectra of this source too, strengthening the conclusion that
these are images of the same galaxy. The WFC3-IR-GRISM data program (\citealt{tre15}) found a redshift of $z=1.24$ for source 13 of J14. Using this information,
we carefully checked our data again, and find evidence for an [\ion{O}{2}] 
emission line very close to severe sky contamination. Due to this contamination,
we are unable to determine which side of the [\ion{O}{2}] doublet we detect, and we set the redshift to
$z=1.240 \pm 0.001$. We do not find any other emission lines in the stacked spectra.

\section{Lens modeling} 
\label{sec:lensmod}
The strong lensing modeling of MACS 1149 has many similarities with the one performed by \citet{gri15} on another HFF cluster, MACS J0416.1$-$2403. The interested reader is referred to that work for a more extensive description of the modeling and statistical analysis. The software used to model MACS 1149 is {\sc Glee}, developed by A. Halkola and S. H. Suyu \citep{SuyuHalkola10,SuyuEtal12}. In the past years, {\sc Glee} has been employed to study the mass distribution of lens galaxies and galaxy clusters, described in terms of physically motivated and simply-parametrized mass profiles, and to probe the expansion history of the Universe through measurements of cosmological parameters (e.g., \citealt{suy13,suy14}).

\subsection{Multiple image systems}
We start with the observed angular positions of several multiple image systems to reconstruct the total gravitational potential of MACS 1149. In detail, we model the positions of 18 and 8 multiple images belonging, respectively, to 7 spectroscopic (6 of which are in our MUSE catalog, see Table \ref{images}) and 3 photometric families of the ``gold'' sample presented in Table 3 of \citet{tre15b}. In Table 4 of the same paper, the approximate coordinates of the 62 multiple images, associated to 18 knots of the Refsdal host, are also listed and are further used as constraints on the mass model. Those are only the ones that were considered very secure by our group. The reliability of the matching of these knots is firmly supported by both the photometric (\HST) and spectroscopic, kinematic, (MUSE) data that will be shown in Figures \ref{fi02} and \ref{fi10}. In summary, the multiple images and knots used in the following constitute the largest subset of those discussed in \citet{tre15b}, of which every team was confident. They were identified iteratively by all the teams during the lens modeling phase and, to this aim, the capabilities of {\sc Glee} have played a decisive role (as further illustrated in Section \ref{subsec:res}).

In Tables \ref{images} and \ref{knots}, we list explicitly the values of the observed angular positions, $x$ and $y$ (measured with respect to the luminosity center of the BCG and positive in the West and North directions), and spectroscopic redshifts, $z_{\rm sp}$, of the 88 multiple images from 11 different background sources considered in our strong lensing models. We observe that the multiple image systems are distributed over a relatively large area of the core of the galaxy cluster and cover a fairly extended range of redshifts (from 1.240 to 3.703). We use the positions of the multiple images as constraints, and the multiple image families provide in total 176 constraints (i.e., $x$ and $y$ coordinates). We note that in our tables we list the previously identified family IDs only, and not the exact image IDs, because of possible small differences in the position values presented in other works. This is due to the fact that we have measured independently the values of $x$ and $y$ from combinations of the available \HST\ images optimized to this purpose. We remark that all multiple images can be well approximated by point-like objects, for which we adopt a constant positional uncertainty of 0.065\arcsec\ (corresponding to one pixel of the chosen \HST\ mosaics). In the following strong lensing analysis, the values of the source redshift of the spectroscopic families will be kept fixed, while those of the photometric families will be optimised in the form of angular-diameter distance ratios $D_{\rm ls}/D_{\rm s}$ with uniform priors on these ratios, where $D_{\rm ls}$ and $D_{\rm s}$ are the lens-source and observer-source angular-diameter distances, respectively.

\begin{table}
\centering
\caption{Angular positions and spectroscopic redshifts of the multiple image systems.}
\begin{tabular}{ccccc}
\hline\hline \noalign{\smallskip}
ID$^{\mathrm{a}}$ & $x^{\mathrm{b}}$ & $y^{\mathrm{b}}$ & $z_{\mathrm{sp}}$ & MUSE ID  \\
 & (arcsec) & (arcsec) & & \\
\noalign{\smallskip} \hline \noalign{\smallskip}
2   & $-$12.26 & $-$31.61 & 1.892 & 112 \\
2   & $-$24.29 & $-$21.85 & 1.892 & 112 \\
2   & $-$26.08 & $-$20.32 & 1.892 & 112 \\
3   &    26.61 &     4.71 & 3.130 & 116 \\
3   &    20.09 &    16.38 & 3.130 & 116 \\
3   &  $-$8.48 &    31.17 & 3.130 & 116 \\
4   &    19.05 &  $-$6.11 & 2.949 & 113 \\
4   &    14.46 &     7.94 & 2.949 & 113 \\
4   & $-$18.16 &    27.32 & 2.949 & 113 \\
5   &  $-$3.35 & $-$19.69 & 2.80  & \\
5   &  $-$7.84 & $-$16.91 & 2.80  & \\
6   &  $-$3.21 & $-$21.55 &       & \\
6   & $-$10.27 & $-$16.82 &       & \\
7   &  $-$0.72 & $-$25.89 &       & \\
7   & $-$15.50 & $-$15.30 &       & \\
7   & $-$29.40 &     9.76 &       & \\
8   &     0.82 & $-$15.05 &       & \\
8   &  $-$3.48 & $-$12.55 &       & \\
8   & $-$27.77 &    22.29 &       & \\ 
13  & $-$16.53 &  $-$2.68 & 1.240 & 106 \\
13  & $-$13.60 &  $-$6.78 & 1.240 & 106 \\
13  &  $-$4.28 & $-$16.82 & 1.240 & 106 \\
14  &    23.55 &    17.93 & 3.703 & 118 \\
14  &    26.31 &    14.80 & 3.703 & 118 \\
110 &  $-$4.68 & $-$30.13 & 3.216 & 117 \\
110 & $-$17.59 & $-$20.26 & 3.216 & 117 \\
\noalign{\smallskip} \hline
\end{tabular}
\begin{list}{}{}
\item[$^{\mathrm{a}}$]Main ID of each family as listed by \citet{tre15b}. Secondary IDs are not matched because of the possible slight coordinate offsets.
\item[$^{\mathrm{b}}$]With respect to the luminosity center of the BCG ($\alpha$~=~11:49:35.699, $\delta$~=~+22:23:54.71) and positive in the West and North directions.
\end{list}
\label{images}
\end{table}

\begin{table}
\centering
\caption{Angular positions of the knots of the `Refsdal' host (MUSE ID 109, $z_{\mathrm{sp}}=1.489$).}
\begin{tabular}{ccc}
\hline\hline \noalign{\smallskip}
ID$^{\mathrm{a}}$ & $x^{\mathrm{b}}$ & $y^{\mathrm{b}}$ \\
 & (arcsec) & (arcsec) \\
\noalign{\smallskip} \hline \noalign{\smallskip}
1.1  &     5.75 &  $-$9.10  \\
1.1  &  $-$3.71 &  $-$5.03  \\
1.1  & $-$15.52 &    14.09  \\
1.1  &  $-$2.24 &  $-$3.93  \\
1.2  &     7.10 &  $-$8.00  \\
1.2  &  $-$0.82 &  $-$2.39  \\
1.2  & $-$14.26 &    14.94  \\
1.2  &     0.26 &  $-$1.04  \\
1.2  &     3.26 &     0.93  \\
1.3  &     6.22 &  $-$8.32  \\
1.3  &  $-$1.40 &  $-$3.35  \\
1.3  & $-$15.09 &    14.63  \\
1.4  &     5.74 &  $-$8.34  \\
1.4  &  $-$1.61 &  $-$3.77  \\
1.4  & $-$15.47 &    14.50  \\
1.5  &     4.95 &  $-$8.36  \\
1.5  &  $-$1.95 &  $-$4.42  \\
1.5  & $-$16.03 &    14.37  \\
1.7  &     3.91 &  $-$8.70  \\
1.7  &  $-$3.30 &  $-$5.75  \\
1.7  & $-$16.49 &    13.95  \\
1.8  &     2.65 &  $-$9.06  \\
1.8  &  $-$3.54 &  $-$6.41  \\
1.8  & $-$16.81 &    13.61  \\
1.10 &     2.17 &  $-$9.63  \\
1.10 &  $-$4.07 &  $-$6.56  \\
1.10 & $-$16.81 &    13.15  \\
1.13 &     5.92 &  $-$6.81  \\
1.13 &  $-$0.27 &  $-$2.95  \\
1.13 & $-$15.13 &    15.47  \\
1.13 &     2.78 &  $-$0.35  \\
1.14 &     5.42 &  $-$6.50  \\
1.14 &  $-$0.10 &  $-$3.23  \\
1.14 & $-$15.44 &    15.63  \\
1.14 &     2.16 &  $-$0.99  \\
1.15 &     5.25 &  $-$7.29  \\
1.15 &  $-$0.75 &  $-$3.80  \\
1.15 & $-$15.63 &    15.20  \\
1.16 &     4.32 &  $-$7.67  \\
1.16 &  $-$1.36 &  $-$4.69  \\
1.16 & $-$16.15 &    14.85  \\
1.17 &     1.99 &  $-$7.86  \\
1.17 &  $-$1.78 &  $-$6.12  \\
1.17 & $-$16.95 &    14.52  \\
1.19 &     6.18 &  $-$9.97  \\
1.19 &  $-$4.07 &  $-$5.13  \\
1.19 & $-$15.38 &    13.54  \\
1.19 &  $-$2.64 &  $-$3.76  \\
1.20 &     5.56 & $-$10.09  \\
1.20 &  $-$4.16 &  $-$5.58  \\
1.20 & $-$15.92 &    13.33  \\
1.20 &  $-$2.94 &  $-$4.21  \\
1.23 &     6.74 & $-$11.34  \\
1.23 &  $-$4.62 &  $-$4.79  \\
1.23 & $-$14.97 &    12.53  \\
1.23 &  $-$3.40 &  $-$3.72  \\
1.24 &     7.47 & $-$10.59  \\
1.24 & $-$14.19 &    13.22  \\
S1   &     1.73 & $-$10.45  \\
S2   &     3.44 &  $-$9.87  \\
S3   &     4.58 & $-$10.76  \\
S4   &     3.15 & $-$12.09  \\
\noalign{\smallskip} \hline
\end{tabular}
\begin{list}{}{}
\item[$^{\mathrm{a}}$]Main ID of each family as listed by \citet{tre15b}. Secondary IDs are not matched because of the possible slight coordinate offsets.
\item[$^{\mathrm{b}}$]With respect to the luminosity center of the BCG ($\alpha$~=~11:49:35.699, $\delta$~=~+22:23:54.71) and positive in the West and North directions.
\end{list}
\label{knots}
\end{table}

\subsection{Mass components}
The strong lensing models considered in this work describe the total mass distribution of MACS 1149 in terms of two different classes of components: cluster members and cluster dark-matter halos. We detail here how many mass components are included in our models and how the total mass density profiles of these components are parametrized in {\sc Glee}.

\begin{figure}
\centering
\includegraphics[width=0.48\textwidth]{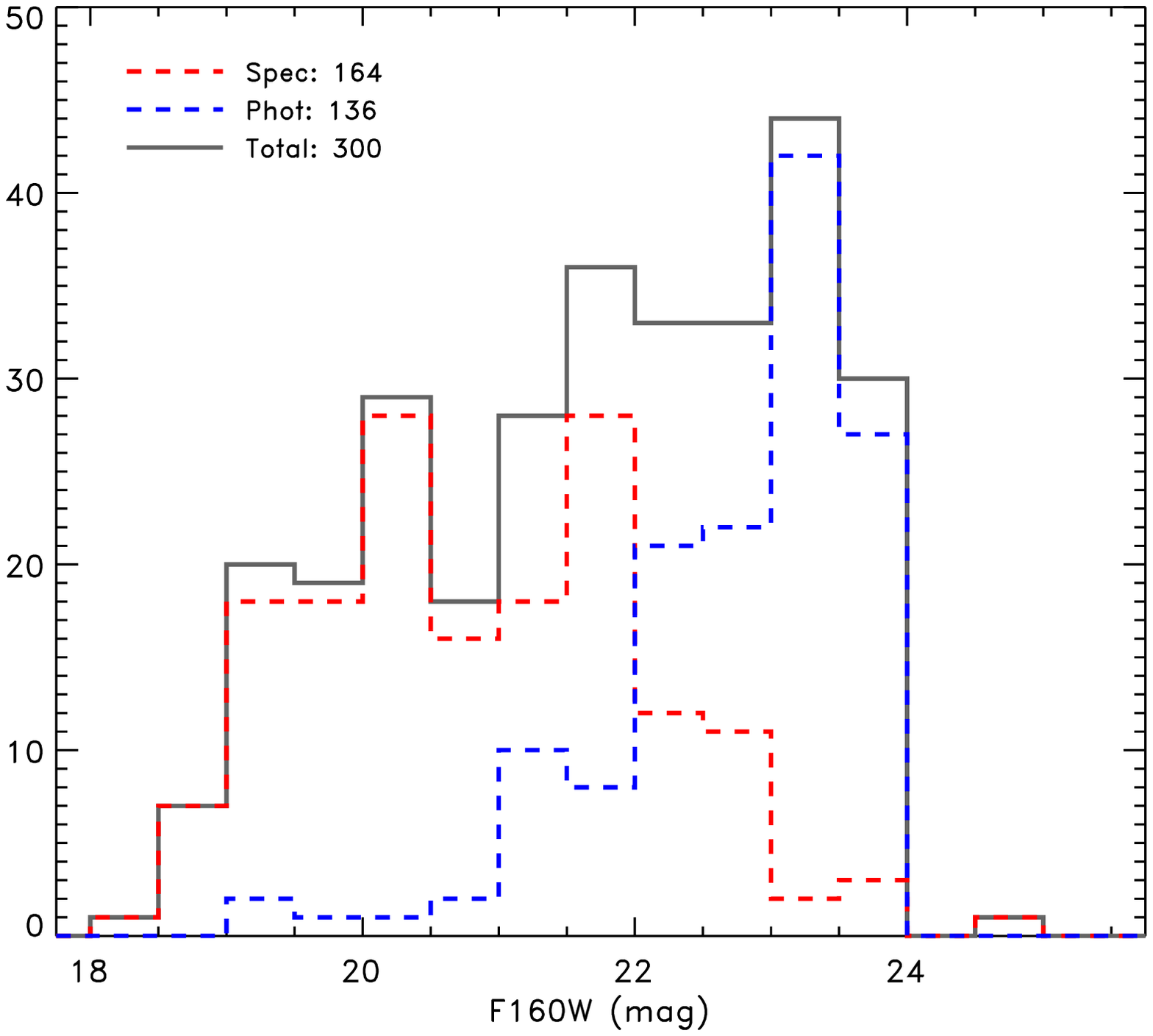}
\caption{Distribution of the Kron magnitudes in the near-IR F160W band of the 300 candidate cluster members included in our strong lensing models. The histograms of the members spectroscopically confirmed and photometrically selected are shown, respectively, in red and blue.}
\label{fi05}
\end{figure}

First, we obtain a spectrophotometric sample of candidate cluster members with the method presented by \citet{gri15}. A brief summary is given below. We start from the CLASH photometric data set in 13 \HST\ broadband filters (i.e., excluding the 3 bluest of the 16 available bands, due to low S/N values of the faint objects) and measure, in each band, the values of the Kron magnitudes (and colors) of all the objects detected in the WFC3/F160W FoV. The Kron magnitudes of each band are extracted in dual mode, always using the combined ACS+WFC3-IR image as detection image. By studying the cumulative luminosity profiles of a large number of objects with different mean surface-brightness, we have checked that the Kron magnitudes well approximate the total model magnitudes (within only a few hundredths of a magnitude) obtained by detailed modeling with the publicly available software {\sc GALFIT} (\citealt{pen02}), thus resulting in reliable and unbiased colors. Using the combined MUSE and WFC3-IR-GRISM redshift catalog (see \citealt{tre15b}), we select from the photometric catalog two subsamples: one containing the objects that we define as spectroscopically confirmed cluster members, i.e.~for which $0.520 < z_{\rm sp} < 0.570$ (164 objects), and one containing the spectroscopic objects that, according to our criteria, do not belong to the galaxy cluster, i.e.~for which $z_{\rm sp} < 0.520$ or $z_{\rm sp} > 0.570$. In passing, we mention that MUSE integral field observations do not target color-selected objects and thus provide a color-unbiased sample of spectroscopic cluster members. Then, for each object, we compare its colors to those of the objects in the previous two subsamples and perform a Bayesian analysis (with all priors chosen uniform) to assign a probability value of being or not being a cluster member. To maximize purity over completeness, in particular at the bright-end of the cluster member luminosity function (i.e., where one finds the most massive galaxies, providing the second most important contribution to a cluster mass model, after the cluster dark-matter halo component), we choose a threshold value of 0.8 on the probability of being a cluster member. In this way, we select a final catalog of candidate cluster members containing 300 objects (see Figure \ref{fi05} and Table \ref{tab10}), down to a magnitude limit of 24 mag in F160W. This corresponds approximately to the magnitude value measured for the faintest spectroscopically confirmed cluster member. We remark that approximately 55\% of the candidate cluster members are spectroscopically confirmed and that the spectroscopic members represent more than 80\% of the candidate members brighter than 22 mag in F160W.

In our strong lensing analysis, the projected dimensionless
surface mass density, a.k.a.~convergence, of the cluster members is modeled 
with a dual pseudoisothermal elliptical mass distribution 
\citep[dPIE;\ ][]{EliasdottirEtal07, SuyuHalkola10} with zero ellipticity and core radius values,
\be
\label{eq:dpie}
\kappa_{\rm g}(x,y) = \frac{\vartheta_{\rm
    E}}{2}\left(\frac{1}{R} -
  \frac{1}{\sqrt{R^2+r_{\rm t}^2}} \right),
\ee
where $(x,y)$ are coordinates and $R$ (=$\sqrt{x^2+y^2})$ is the 
radial coordinate on the image plane, $\vartheta_{\rm E}$ is the galaxy lens strength
(a.k.a.~Einstein radius), and $r_{\rm t}$ is the
``truncation radius''. A dPIE mass distribution is centered on the luminosity centroid position ($x_{\rm g}$, $y_{\rm g}$) of each cluster member (see Table \ref{tab10}). The three-dimensional mass density
distribution corresponding to Equation (\ref{eq:dpie}) is
\be
\label{eq:dpie_rho}
\rho(r) \propto \frac{1}{r^2(r^2+r_{\rm t}^2)},
\ee
where $r$ is the three dimensional radius ($r=\sqrt{x^2+y^2+z^2}$).
This mass density distribution is ``isothermal'', i.e. it scales as 
$r^{-2}$, in the core ($r \ll r_{\rm t}$) and is ``truncated'', 
i.e. it scales as $r^{-4}$, in the outer regions ($r \gg r_{\rm t}$).
The parameter $r_{\rm t}$ corresponds approximately to
the half-mass radius \citep[e.g.,][]{EliasdottirEtal07}.

Two spectroscopically confirmed cluster members, labeled as G1 and G2 in Table \ref{tab10}, are angularly very close to the Refsdal multiple images S1-S4 and SX for which we want to determine accurate magnification and time-delay values. To allow for more flexibility in our strong lensing models, we also consider elliptical dPIE components with axis ratio and position angles as free parameters for these two galaxies (where we replace $R$ in Equation (1) by 
\be
\label{eq:rem}
R_{\rm \epsilon}^2 =\frac{{x^2}}{(1+\epsilon)^2}+\frac{y^2}{(1-\epsilon)^2}
\ee
with the ellipticity $\epsilon$ defined in terms of the axis ratio $q$ as $\epsilon=(1-q)/(1+q)$, and allow the $\kappa_{\rm g}$ distribution to be rotated by the position angle $\phi$). Moreover, to test the importance of the choice of the radial slope value of the total mass distribution, G1 and G2 can also be modeled as singular power-law ellipsoids \citep[SPLE;\ ][]{bar98} with dimensionless surface mass density
\be
\label{eq:spemd}
\kappa(x,y) = b \left(x^2+\frac{y^2}{q^2}\right)^{(1-\gamma')/2},
\ee
where $b$ is the galaxy lens strength determining the value of the Einstein radius, $q$ is the axis ratio, and $\gamma'$ is the radial slope. The distribution is then rotated by the position angle parameter $\phi$. The ``isothermal'' approximation is obtained when $\gamma'$ is equal to 2.

On radial scales larger than those of the cluster members, we parametrize the total mass distribution of the galaxy cluster with two-dimensional, pseudo-isothermal elliptical mass distributions (PIEMD; \citealt{kas93}). Following the results of previous strong lensing analyses of MACS 1149 (e.g., \citealt{smi09}; \citealt{rau14}; J14) and given its complex merging state, as revealed by deep \emph{Chandra} X-ray observations (Ogrean et al., submitted), we include three such mass components to represent the contribution to the cluster total mass of all remaining mass not associated to the cluster members. 
Since these three PIEMDs are
mainly in the form of dark matter, we will refer to these components as cluster dark-matter halos. The dimensionless surface mass density of a PIEMD profile is 
\be
\label{eq:piemd}
\kappa_{\rm h}(x,y) = \frac{\vartheta_{\rm E,h}}{2\sqrt{R_{\rm \epsilon}^2 + r_{\rm c,h}^2}},
\ee
where $R_{\rm \epsilon}$ is defined in Equation (\ref{eq:rem}) and $q_{\rm h}$ is the dark matter halo axis ratio. The halo strength is $\vartheta_{\rm E,h}$ and the distribution has a
central core radius $r_{\rm c,h}$ that defines where the radial dependence of $\kappa_{\rm h}$ changes from $R^0$ to $R^{-1}$. The distribution can be shifted to the mass center of 
each
 cluster
halo ($x_{\rm h}$, $y_{\rm h}$) and rotated by its
position angle $\phi_{\rm h}$. Each PIEMD is thus characterized by 6 parameters ($x_{\rm h}$, $y_{\rm h}$, $q_{\rm h}$, $\phi_{\rm  h}$, $\vartheta_{\rm E,h}$, $r_{\rm c,h}$).

\subsection{Mass models}
We investigate several mass models, each comprised of cluster galaxies and cluster dark-matter 
halos, to find the one that best reproduces the observed positions of the multiple images of Tables \ref{images} and \ref{knots} and to estimate the systematic uncertainties on the magnification and time-delay values at the locations of the Refsdal multiple images.

We include the mass contribution of the 300 cluster members identified above and modeled with dPIE profiles. We consider the galaxy luminosity values, $L$, in the \HST\  F160W band (i.e., the reddest of the available near-IR bands) to ``weight'' their relative total mass. In detail, we adopt the following two scaling relations for the values of the Einstein radius, $\vartheta_{\mathrm{E},i}$, and truncation radius, $r_{\mathrm{t},i}$, of the $i$-th cluster member:
\begin{equation}
\label{eq:sr1}
\vartheta_{\mathrm{E},i}=\vartheta_{\mathrm{E,g}}
\left(\frac{L_{i}}{L_{\rm g}}\right)^{0.5} \quad \textrm{and} \quad
r_{\mathrm{t},i}=r_{\mathrm{t,g}} \left(\frac{L_{i}}{L_{\rm g}}\right)^{0.5} \, 
\end{equation}
and
\begin{equation}
\label{eq:sr2}
\vartheta_{\mathrm{E},i}=\vartheta_{\mathrm{E,g}}
\left(\frac{L_{i}}{L_{\rm g}}\right)^{0.7} \quad \textrm{and} \quad
r_{\mathrm{t},i}=r_{\mathrm{t,g}} \left(\frac{L_{i}}{L_{\rm g}}\right)^{0.5} \, ,
\end{equation}
where $\vartheta_{\mathrm{E},g}$ and $r_{\mathrm{t},g}$ are two reference values, chosen, in our models, equal to those of the second brightest cluster galaxy, i.e., ``Ref'' in Table \ref{tab10}.
 Given the relation between the value of total mass, $M_{\mathrm{T}}$, and those of effective velocity dispersion, $\sigma$ (where $\sigma \sim \vartheta_{\mathrm{E}}^{0.5}$), and truncation radius, $r_{\rm t}$, for a dPIE profile, it follows that the previous two relations correspond, respectively, to
\begin{equation}
\label{eq:mr1}
\frac{M_{\mathrm{T},i}}{L_{i}} \sim \frac{\sigma_{i}^{2}r_{\mathrm{t},i}}{L_{i}} \sim \frac{L_{i}^{0.5}L_{i}^{0.5}}{L_{i}} \sim L_{i}^{0} \, 
\end{equation}
and
\begin{equation}
\label{eq:mr2}
\frac{M_{\mathrm{T},i}}{L_{i}} \sim \frac{\sigma_{i}^{2}r_{\mathrm{t},i}}{L_{i}} \sim \frac{L_{i}^{0.7}L_{i}^{0.5}}{L_{i}} \sim L_{i}^{0.2}.
\end{equation}
The relations in Equations (\ref{eq:sr1}) and (\ref{eq:sr2}) are therefore equivalent to having cluster members with total mass-to-light ratios that are constant ($M_{\mathrm{T}}L^{-1}=k$) and that increase with the luminosity ($M_{\mathrm{T}}L^{-1}\sim L^{0.2}$), respectively. The last relation between $M_{\mathrm{T}}/L$ and $L$ is also known as the tilt of the Fundamental Plane (e.g., \citealt{fab87}; \citealt{ben92}) and is used to describe the systematic increase of galaxy effective mass-to-light ratio with effective mass measured in early-type galaxies.  Therefore, for the mass distribution associated with all the cluster members, we have effectively 2 free parameters: $\vartheta_{\mathrm{E,g}}$ and $r_{\mathrm{t,g}}$.

As mentioned in the previous section, all models contain three extended and smooth dark-matter halo components, parametrized by three PIEMD profiles. The centers of the three components are initially on (1) the cluster BCG, (2) the second brightest cluster galaxy (labeled as ``Ref'' in Table \ref{tab10}), and (3) a dense group of cluster members, approximately 50 arcsec to the North-West of the BCG. The values of the 18 parameters ($=6\times3$) describing these three mass components have broad uniform priors and are all optimized.

The combination of the aforementioned cluster dark-matter halos and cluster galaxies yields two different models: (1) 3PIEMD + 300dPIE ($M_{\mathrm{T}}L^{-1}=k$) or ``MLC G12L'', and (2) 3PIEMD + 300dPIE ($M_{\mathrm{T}}L^{-1}\sim L^{0.2}$) or ``MLV G12L''. The cluster total mass distribution of both models is characterized by 20 free parameters: 18 for the cluster dark-matter halos and 2 for the cluster members (all, including G1 and G2, approximated by circular total mass profiles).

In addition, we try four models with more flexible mass distributions for the two cluster members G1 and G2 (in projection, the closest galaxies to the Refsdal multiple images S1-S4 and SX), keeping all other cluster members and the cluster dark-matter halos the same parameterization as before.  In the first set of two models, G1 and G2 are represented with two independent dPIE profiles with their values of lens strength, truncation radius, axis ratio and position angle free to vary (i.e., no longer following the $M_{\mathrm{T}}/L$ scaling relations); these two models are identified with 3PIEMD + 298dPIE ($M_{\mathrm{T}}L^{-1}=k$) + 2dPIE or ``MLC G12F'', and 3PIEMD + 298dPIE ($M_{\mathrm{T}}L^{-1}\sim L^{0.2}$) + 2dPIE or ``MLV G12F''.  In the second set of two models, G1 and G2 are represented instead with two independent SPLE profiles with their values of lens strength, radial slope, axis ratio and position angle free to vary; the final two models are denoted by 3PIEMD + 298dPIE ($M_{\mathrm{T}}L^{-1}=k$) + 2SPLE or ``MLC G12P'' and 3PIEMD + 298dPIE ($M_{\mathrm{T}}L^{-1}\sim L^{0.2}$) + 2SPLE or ``MLV G12P''.  The cluster total mass distribution of these four additional models is characterized by 28 free parameters: 18 for the cluster dark-matter halos, 2 for the cluster members and 8 for G1 and G2.

\subsection{Results}
\label{subsec:res}

\begin{table*}
\centering
\caption{The investigated strong lensing models and their values of 
best-fitting $\chi^2$ (mininum-$\chi^2$), degrees of freedom (dof), 
 and rms offset between observed and model-predicted multiple image positions.}
\begin{tabular}{ccccc}
\hline\hline \noalign{\smallskip}
ID & Model & $\chi^{2}$ & dof & rms \\
 & & & & (arcsec) \\
\noalign{\smallskip} \hline \noalign{\smallskip}
MLC G12L & 3PIEMD + 300dPIE ($M_{\mathrm{T}}L^{-1}=k$) & 5094 & 97 & 0.49 \\
MLV G12L & 3PIEMD + 300dPIE ($M_{\mathrm{T}}L^{-1}\sim L^{0.2}$) & 5039 & 97 & 0.49 \\
MLC G12F & 3PIEMD + 298dPIE ($M_{\mathrm{T}}L^{-1}=k$) + 2dPIE & 1511 & 89 & 0.27 \\
MLV G12F & 3PIEMD + 298dPIE ($M_{\mathrm{T}}L^{-1}\sim L^{0.2}$) + 2dPIE & 1441 & 89 & 0.26 \\
MLC G12P & 3PIEMD + 298dPIE ($M_{\mathrm{T}}L^{-1}=k$) + 2SPLE & 3324 & 89 & 0.40 \\
MLV G12P & 3PIEMD + 298dPIE ($M_{\mathrm{T}}L^{-1}\sim L^{0.2}$) + 2SPLE & 3335 & 89 & 0.40 \\
\noalign{\smallskip} \hline
\end{tabular}
\label{models}
\end{table*}

We summarize in Table \ref{models} the values of the 
best-fitting $\chi^2$ (i.e., minimum $\chi^2$) 
and root mean square (rms) offset between the observed and model-predicted positions of the multiple images for the six different cluster mass models described above. We find that the models that best reproduce the strong lensing observables are those in which G1 and G2 are parametrized with dPIE profiles with their parameter values free to vary, followed by the models where G1 and G2 are described by SPLE profiles, and finally by those in which G1 and G2 are represented by dPIE profiles with parameter values linked to those of the other clusters members, according to the Equations (\ref{eq:sr1}) and (\ref{eq:sr2}). The absolute best-fitting mass model (MLV G12F) contains 3 extended dark-matter halo components in the form of cored elliptical pseudo-isothermal mass distributions and 300 candidate cluster members modeled as dual pseudo-isothermal mass distributions; of the 300 candidate cluster members, 298 are approximated as axially symmetric and scaled with total mass-to-light ratios increasing with their near-IR luminosities, and 2 are elliptical with 
mass parameters free to vary. This model provides a minimum $\chi^{2}$ value of 1441 and can reproduce very accurately the observed positions of the considered 88 multiple images, with a rms offset of only 0.26\arcsec. The results of this particular model are the ones used in the comparison by \citet{tre15b}.

We sample the posterior probability distribution function of the parameters of the MLV G12F lensing model using a standard Bayesian analysis and Markov chain Monte Carlo (MCMC) methods (for more details, see Sect.~3.2 in \citealt{gri15}). We increase the positional errors of the observed multiple images to 0.26\arcsec\  to get a $\chi^{2}$ value (90) that is comparable to the number (89) of the degrees of freedom (dof). The latter is given by the difference between the number of lensing observables (176 $x$ and $y$ coordinates of the multiple images) and that of the model free parameters (28 describing the cluster total mass distribution, 56 $x$ and $y$ coordinates of lensed sources and 3 redshifts of the photometric families). In this way, possible line-of-sight mass structures, small dark-matter substructures, deviations from elliptical mass profiles and some scatter in the adopted scaling relations for the cluster members, which have not been explicitly included in our model, are statistically taken into account, and realistic errors on the values of the model parameters can be estimated. We obtain a final MCMC chain with $10^{6}$ samples with an acceptance rate of approximately 0.13.

\begin{table*}
\centering
\caption{Model-predicted values of position, magnification, and time delay of the multiple images of Refsdal.}
\begin{tabular}{ccccccccc}
\hline\hline \noalign{\smallskip}
 & & MLC G12L & MLV G12L & MLC G12F & MLV G12F$^{\dagger}$ & MLC G12P & MLV G12P &  MLV G12F$^{\mathrm{c}}$ \\
\noalign{\smallskip} \hline \noalign{\smallskip}
$x_{\rm S1}$$^{\mathrm{a}}$ & (arcsec) & 1.97 & 2.01 & 1.71 & 1.75 & 1.80 & 1.80 & 1.75 [1.63,1.90] \\
$y_{\rm S1}$$^{\mathrm{a}}$ & (arcsec) & $-$10.46 & $-$10.40 & $-$10.54 & $-$10.50 & $-$10.47 & $-$10.40 & $-$10.42 [$-$10.48,$-$10.37] \\
$\mu_{\rm S1}$ & & 23.3 & 25.2 & 14.3 & 16.0 & 13.1 & 12.0 & 13.5 [10.3,17.4] \\
$\Delta t_{\rm S1}$$^{\mathrm{b}}$ & (days) & $\equiv$0.0 & $\equiv$0.0 & $\equiv$0.0 & $\equiv$0.0 & $\equiv$0.0 & $\equiv$0.0 & $\equiv$0.0 $\equiv$[0.0,0.0] \\
\noalign{\smallskip} \hline \noalign{\smallskip}
$x_{\rm S2}$$^{\mathrm{a}}$ & (arcsec) & 3.23 & 3.26 & 3.28 & 3.29 & 3.34 & 3.44 & 3.40 [3.32,3.54] \\
$y_{\rm S2}$$^{\mathrm{a}}$ & (arcsec) & $-$10.04 & $-$10.03 & $-$9.94 & $-$9.97 & $-$9.94 & $-$9.93 & $-$9.95 [$-$10.03,$-$9.85] \\
$\mu_{\rm S2}$ & & $-$25.3 & $-$29.1 & $-$12.6 & $-$14.3 & $-$12.6 & $-$11.96 & $-$12.4 [$-$18.8,$-$7.9] \\
$\Delta t_{\rm S2}$$^{\mathrm{b}}$ & (days) & 3.9 & 3.5 & 10.6 & 9.4 & 9.1 & 10.1 & 10.6 [7.6,16.8] \\
\noalign{\smallskip} \hline \noalign{\smallskip}
$x_{\rm S3}$$^{\mathrm{a}}$ & (arcsec) & 4.27 & 4.25 & 4.51 & 4.49 & 4.47 & 4.51 & 4.53 [4.41,4.61] \\
$y_{\rm S3}$$^{\mathrm{a}}$ & (arcsec) & $-$10.51 & $-$10.51 & $-$10.67 & $-$10.72 & $-$10.66 & $-$10.74 & $-$10.79 [$-$10.88,$-$10.66] \\
$\mu_{\rm S3}$ & & 22.4 & 25.0 & 13.9 & 15.2 & 13.8 & 13.3 & 13.4 [10.3,19.2] \\
$\Delta t_{\rm S3}$$^{\mathrm{b}}$ & (days) & 1.3 & 1.3 & 3.8 & 3.2 & 4.2 & 5.2 & 4.8 [3.0,8.0] \\
\noalign{\smallskip} \hline \noalign{\smallskip}
$x_{\rm S4}$$^{\mathrm{a}}$ & (arcsec) & 3.15 & 3.12 & 3.13 & 3.11 & 3.11 & 3.05 & 3.05 [2.96,3.10] \\
$y_{\rm S4}$$^{\mathrm{a}}$ & (arcsec) & $-$11.74 & $-$11.75 & $-$11.95 & $-$11.90 & $-$11.97 & $-$12.00 & $-$11.97 [$-$12.11,$-$11.83] \\
$\mu_{\rm S4}$ & & $-$7.0 & $-$7.7 & $-$5.9 & $-$6.4 & $-$4.6 & $-$4.3 & $-$5.7 [$-$8.1,$-$3.7] \\
$\Delta t_{\rm S4}$$^{\mathrm{b}}$ & (days) & 19.6 & 19.0 & 25.9 & 24.0 & 27.5 & 29.2 & 25.9 [21.6,34.0] \\
\noalign{\smallskip} \hline \noalign{\smallskip}
$x_{\rm SX}$$^{\mathrm{a}}$ & (arcsec) & $-$4.68 & $-$4.48 & $-$4.70 & $-$4.61 & $-$4.64 & $-$4.59 & $-$4.59 [$-$4.66,$-$4.50] \\
$y_{\rm SX}$$^{\mathrm{a}}$ & (arcsec) & $-$6.72 & $-$6.73 & $-$6.58 & $-$6.52 & $-$6.45 & $-$6.51 & $-$6.59 [$-$6.66,$-$6.52] \\
$\mu_{\rm SX}$ & & $-$5.0 & $-$5.1 & $-$5.3 & $-$5.2 & $-$5.2 & $-$4.7 & $-$4.8 [$-$5.3,$-$4.5] \\
$\Delta t_{\rm SX}$$^{\mathrm{b}}$ & (days) & 348 & 328 & 366 & 353 & 387 & 390 & 361 [334,381] \\
\noalign{\smallskip} \hline \noalign{\smallskip}
$x_{\rm SY}$$^{\mathrm{a}}$ & (arcsec) & $-$17.03 & $-$16.93 & $-$16.90 & $-$16.92 & $-$16.87 & $-$16.78 & $-$16.87 [$-$16.96,$-$16.82] \\
$y_{\rm SY}$$^{\mathrm{a}}$ & (arcsec) & 12.46 & 12.66 & 12.60 & 12.68 & 12.55 & 12.65 & 12.67 [12.58,12.77] \\
$\mu_{\rm SY}$ & & 5.0 & 4.3 & 4.5 & 4.1 & 4.1 & 3.5 & 4.0 [3.8,4.2] \\
$\Delta t_{\rm SY}$$^{\mathrm{b}}$ & (days) & $-$6017 & $-$6132 & $-$5993 & $-$6118 & $-$6165 & $-$6457 & $-$6183 [$-$6327,$-$6023] \\
\noalign{\smallskip} \hline
\end{tabular}
\begin{list}{}{}
\item[$^{\dagger}$]Reference model.
\item[$^{\mathrm{a}}$]With respect to the luminosity center of the BCG ($\alpha =$ 11:49:35.699, $\delta =$ +22:23:54.71) and positive in the West and North directions.
\item[$^{\mathrm{b}}$]With respect to the peak brightness of S1 which occured on 2015 April 26 ($\pm$20 days in the observer frame; version 1 of \citealt{tre15b}). We note that a slightly different estimate of this epoch has been discussed in detail by \citet{rod16}, after the completion of this work.
\item[$^{\mathrm{c}}$]Median values and 68\% confidence level intervals from the MCMC chain.
\end{list}
\label{stat}
\end{table*}

\begin{figure*}
\centering
\includegraphics[width=0.7\textwidth]{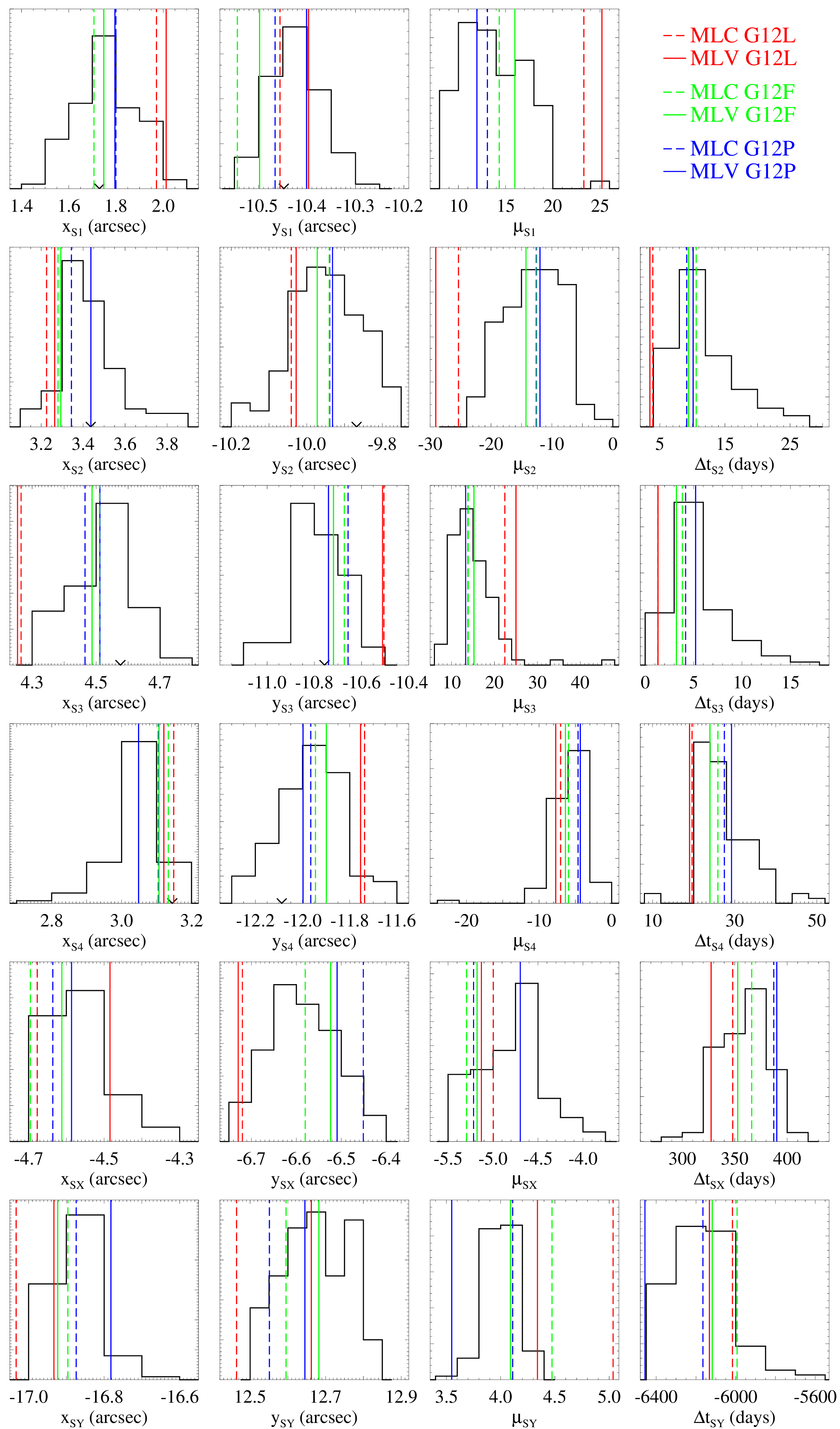}
\caption{Position, magnification, and time delay of the recently observed (S1-S4), future (SX), and past (SY) multiple images of Refsdal. The predicted values of the best-fitting model for each of the six model parameterizations listed in Table \ref{models} are shown with colored solid and dashed lines, according to the legend on the top right. The histograms represent the probability distribution function of the same quantities obtained from 100 different models extracted from the MCMC chain of the best-fitting model (MLV G12F). The V-shaped tick marks in the left two columns for S1-S4 indicate the observed positions of the Refsdal images. According to the value of $\Delta t_{\rm SX}$ and the measured peak luminosity of S1 (see \citealt{tre15b}), SX is predicted to occur between March and June 2016.}
\label{fi01}
\end{figure*}

We show in Table \ref{stat} and Figure \ref{fi01} the model-predicted values of position ($x$ and $y$), magnification $\mu$, and time delay $\Delta t$ of the four recently observed (S1-S4), the future (SX), and the past (SY) images of Refsdal. We provide the best-fitting values obtained from the optimized six lensing models of Table \ref{models} and the 68\% confidence level intervals and probability distribution functions estimated from 100 different models belonging to the final MCMC chain of the MLV G12F model. 

We notice that the systematic errors, associated with the different lensing models considered in this work, are in general slightly larger than the statistical errors, derived from our MCMC analysis. We remark that the models G12F and G12P (plotted in green and blue in Figure \ref{fi01}), in which the two cluster members closest in projection to the identified knots of the Refsdal host have the most flexible total mass profiles, match the positions of the observed multiple images S1-S4 the best (and in general those of all observed multiple images, as indicated by the lower rms values in Table \ref{models}). For the quantities relative to the past and future images of Refsdal, these models also predict values that are in general more consistent than those given by the models G12L (plotted in red in Figure \ref{fi01}), in which the total mass profiles of G1 and G2 are scaled together with those of all other cluster members.

In Figures 9 and 10 of \citet{tre15b}, the good agreement between the observed and reconstructed values of magnification ratio and time delay of the four Refsdal multiple images S1-S4 corroborates the goodness of our MLV G12F model. For image SX, the image that is predicted to appear next, we find that the statistical uncertainty on the absolute value of the magnification factor $\mu_{\rm SX}=4.8^{+0.5}_{-0.3}$ is comparable to the systematic uncertainty (where values range from 4.7 to 5.3 across the different models). According to our best-fitting model, the next image SX should be approximately 20\% fainter than S4 (the least luminous of the four previously observed images of Refsdal), and should be detectable with the planned \HST\ observations. For the value of the time delay of SX with respect to S1, we also obtain that the statistical ($361^{+20}_{-27}$ days) and systematic (from 328 to 390 days) uncertainties are very similar. Given the estimated occurrence (see version 1 of \citealt{tre15b}) of the brightness peak of S1 on April 26 2015 (with an uncertainty of 20 days), our MLV G12F model predicts that the brightness peak of SX should be measured approximately between the middle of March 2016 and the middle of June 2016, and that the possible first detection of SX should be towards the end of 2015.

\begin{figure*}
\centering
\includegraphics[width=0.3\textwidth]{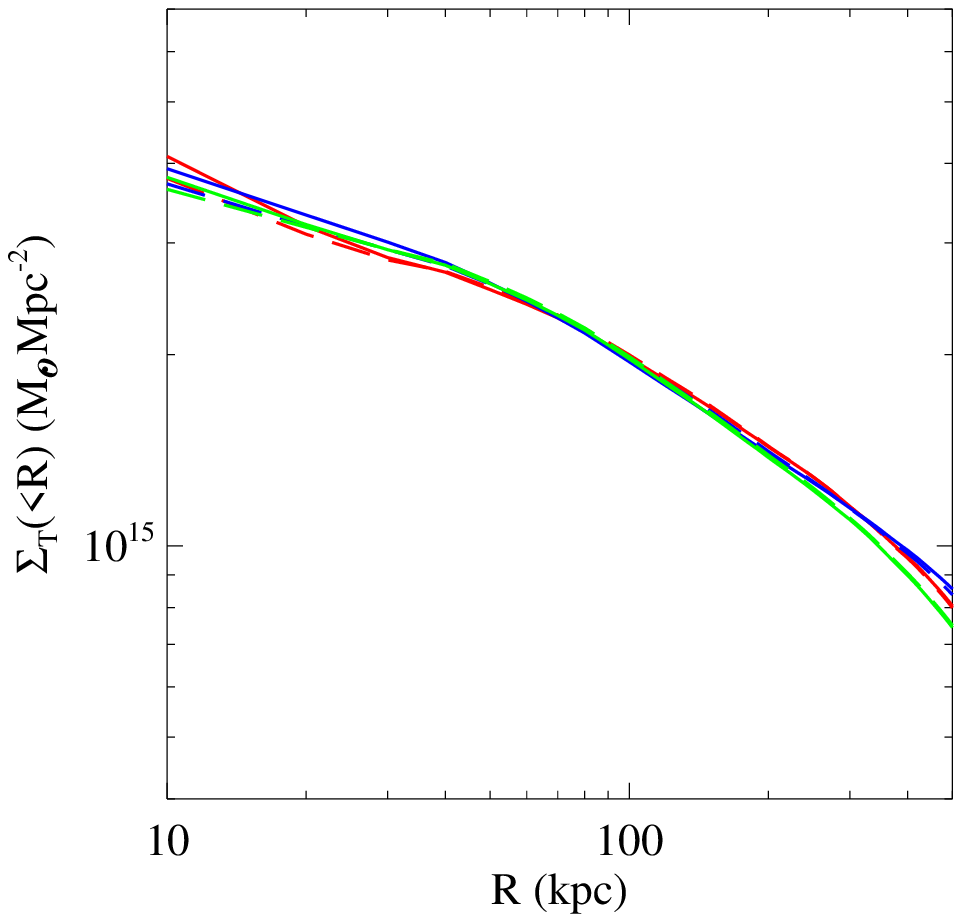}
\includegraphics[width=0.3\textwidth]{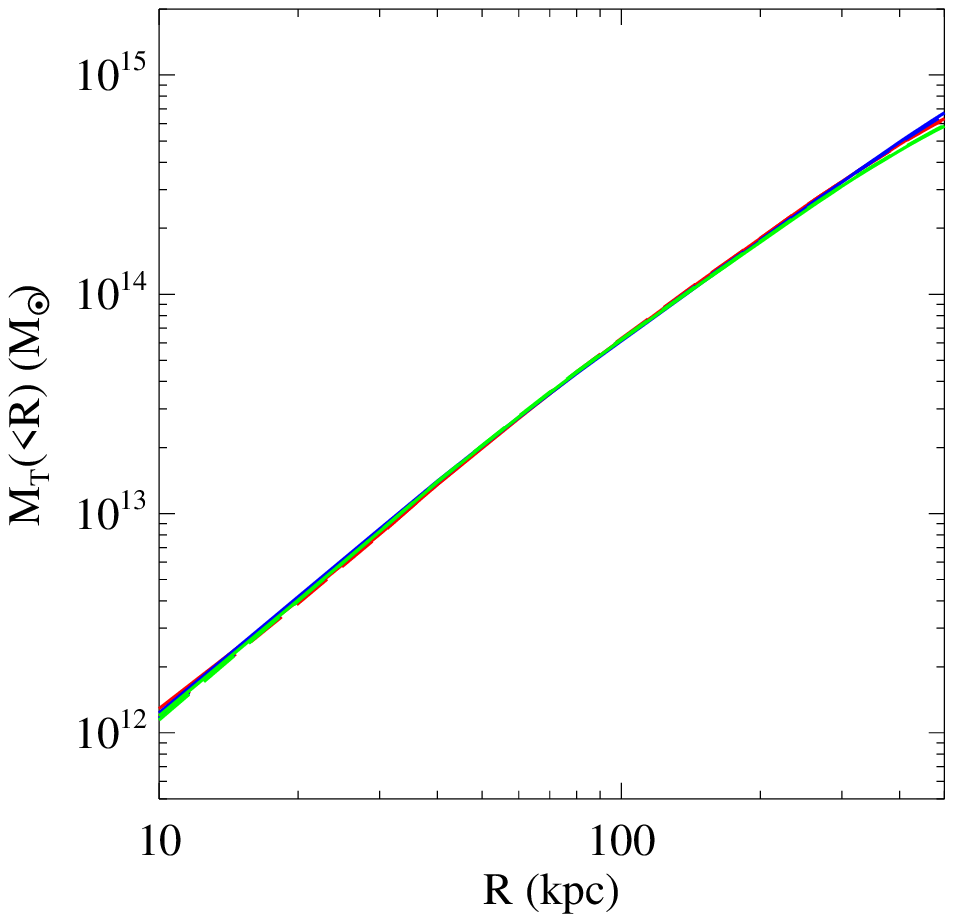}\\
\includegraphics[width=0.3\textwidth]{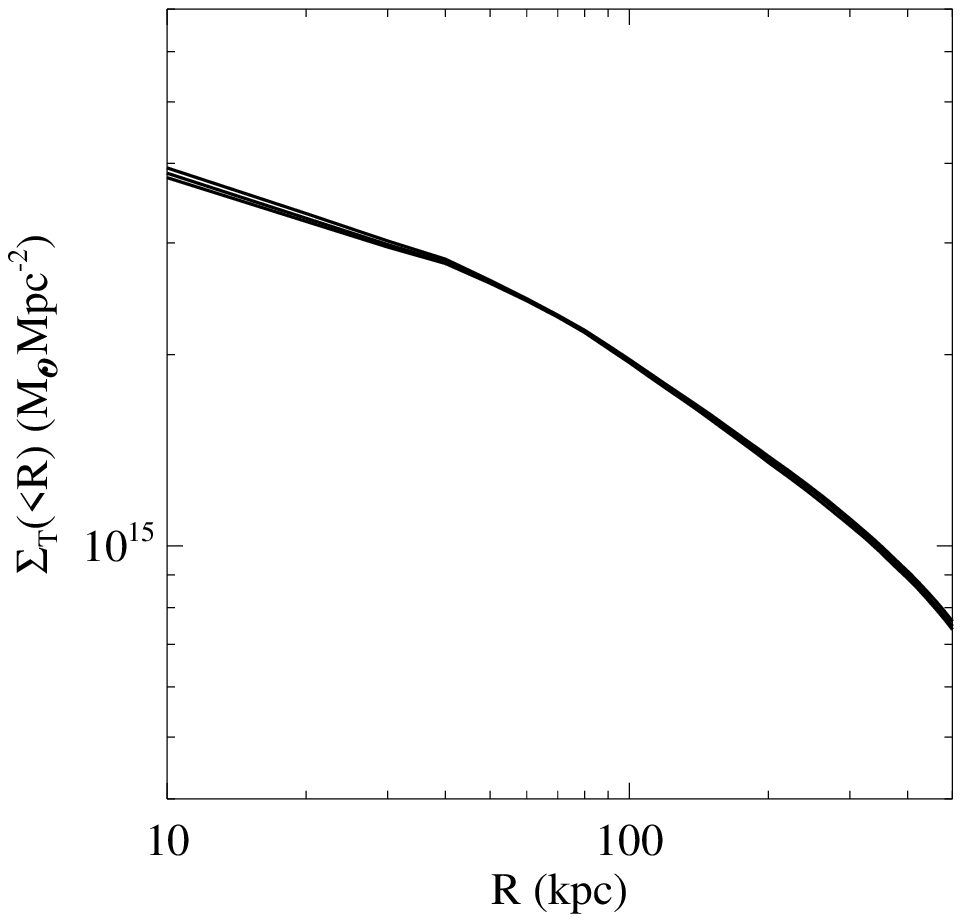}
\includegraphics[width=0.3\textwidth]{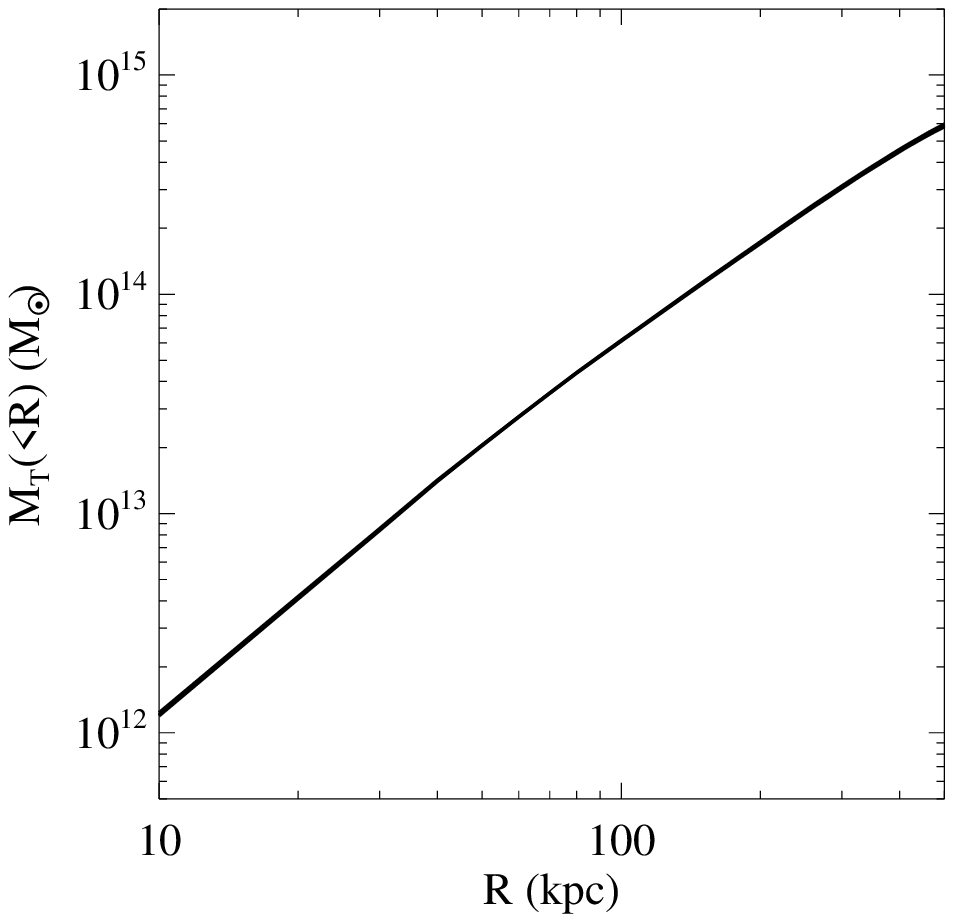}
\caption{Total surface mass density, $\Sigma_{\rm T}(<R)$, and cumulative projected mass, $M_{\rm T}(<R)$, profiles of the six optimized strong lensing models of Table \ref{models} (\emph{on the top}), color coded as in Figure \ref{fi01}, and of the median, with 68\% confidence level uncertainties, MLV G12F model (\emph{on the bottom}), as obtained from the MCMC analysis.}
\label{fi03}
\end{figure*}

We illustrate in Figure \ref{fi03} the total surface mass density profile and enclosed mass of our models.  The circularized total average surface mass density is computed via 
\begin{equation}
\label{eq:asmd}
\Sigma_{\rm T}(<R) \equiv \frac{\int^{R}_{0} \Sigma_{\rm T}(\tilde{R})2\pi\tilde{R}\,\mathrm{d}\tilde{R}}{\pi R^2}
\end{equation}
and cumulative projected mass is
\begin{equation}
\label{eq:cpm}
M_{\rm T}(<R) \equiv \int^{R}_{0} \Sigma_{\rm T}(\tilde{R})2\pi\tilde{R}\,\mathrm{d}\tilde{R} \, ,
\end{equation}
where $\boldsymbol{\tilde{R}} = \tilde{R} \,\boldsymbol{e_{\tilde{R}}} = (x,y)$ and $ \boldsymbol{e_{\tilde{R}}} = \boldsymbol{\tilde{R}} / \tilde{R}$. Although MACS 1149 is a complex merging cluster, we have checked that its barycenter, or center of mass (see Equation (21) in \citealt{gri15}), reconstructed from our optimized strong lensing models is on average only about 10 kpc away in projection from the BCG center. For this reason and to perform a simple comparison of the different models, we approximate the cluster barycenter with the BCG center and measure, on the lens plane, the distances $R$ from the luminosity center of the BCG (see Table \ref{tab10}). 

As visible in Figure \ref{fi03}, our best-fitting MLV G12F model has a relatively flat profile of $\Sigma_{\rm T}(<R)$ in the cluster core, with an average logarithmic slope value of $-0.30 \pm 0.01$ in the range 10-100 kpc, and a total mass value of $(5.9 \pm 0.1) \times 10^{14}$ M$_{\odot}$ projected within a cylinder with radius equal to 500 kpc. Between 10 and 500 kpc in projection from the cluster center, we find that for $\Sigma_{\rm T}(<R)$ and $M_{\rm T}(<R)$ the statistical relative errors, derived from the 100 different models extracted from the MCMC chain, are on the order of only a few percent and the systematic relative errors, estimated from the six different cluster total mass models, reach a maximum value of 8\% in the first radial bin of 10 kpc. The statistical and systematic uncertainties of these quantities show a minimum at approximately 70 kpc from the BCG center. This is not surprising since it is at this radial range 
of the cluster that the majority of the strong lensing information (i.e., the positions of knots of the Refsdal host) is concentrated. 

\begin{figure*}
\centering
\includegraphics[width=0.8\textwidth]{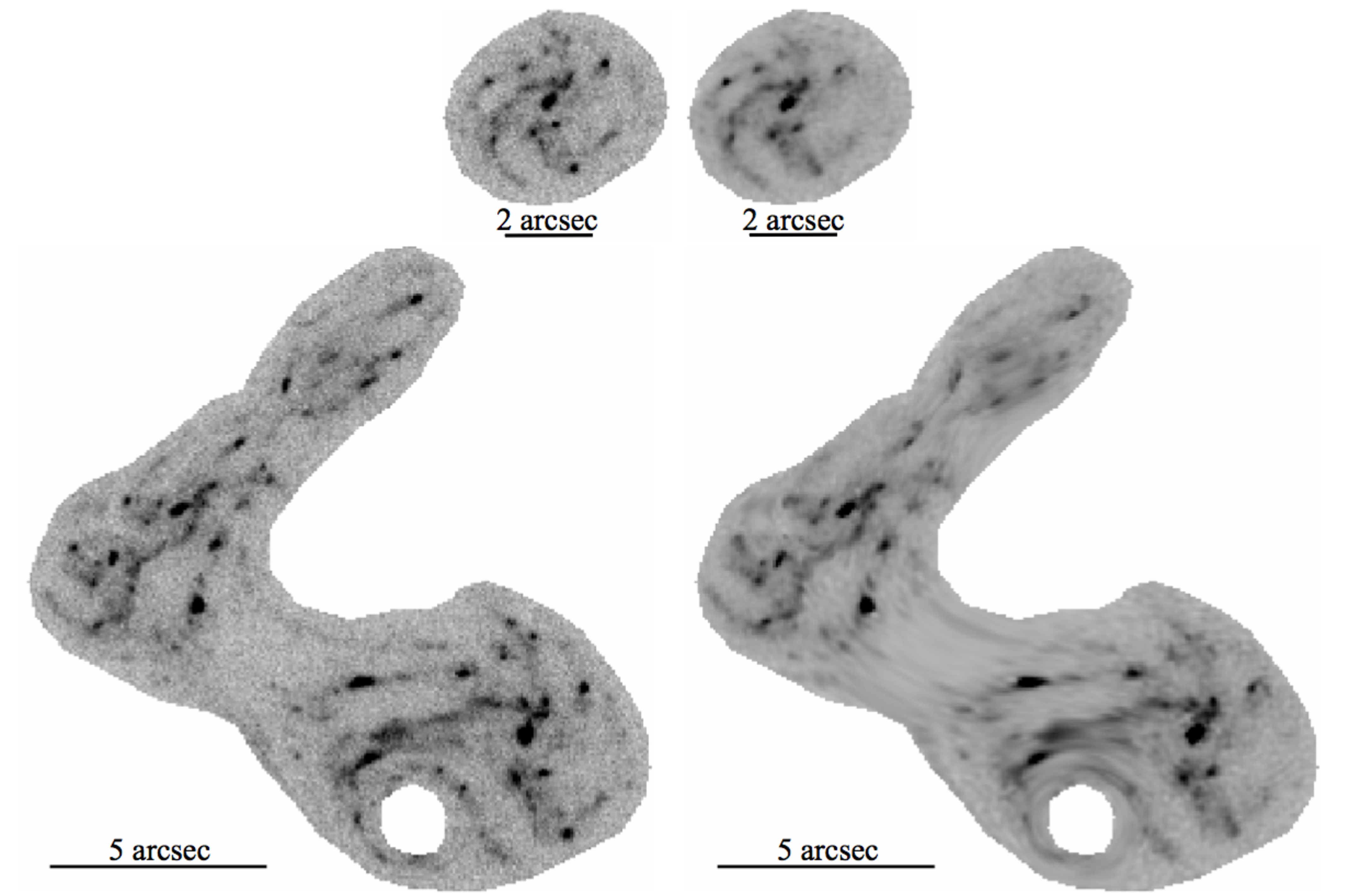}
\caption{Observed (\emph{on the left}) and model-predicted (MLV G12F; \emph{on the right}) surface brightness distribution of the Refsdal host. The past (SY), future (SX), and recently observed (S1-S4) multiple images of Refsdal are located in the top, middle, and bottom images of the host, respectively. The original data is a combined mosaic, with a pixel size of 0.06\arcsec, of the \HST\ F606W and F435W bands from the HFF project (data release v1.0), optimized to suppress the flux contamination by the cluster member galaxies.}
\label{fi02}
\end{figure*}

\begin{figure}
\centering
\includegraphics[width=0.48\textwidth]{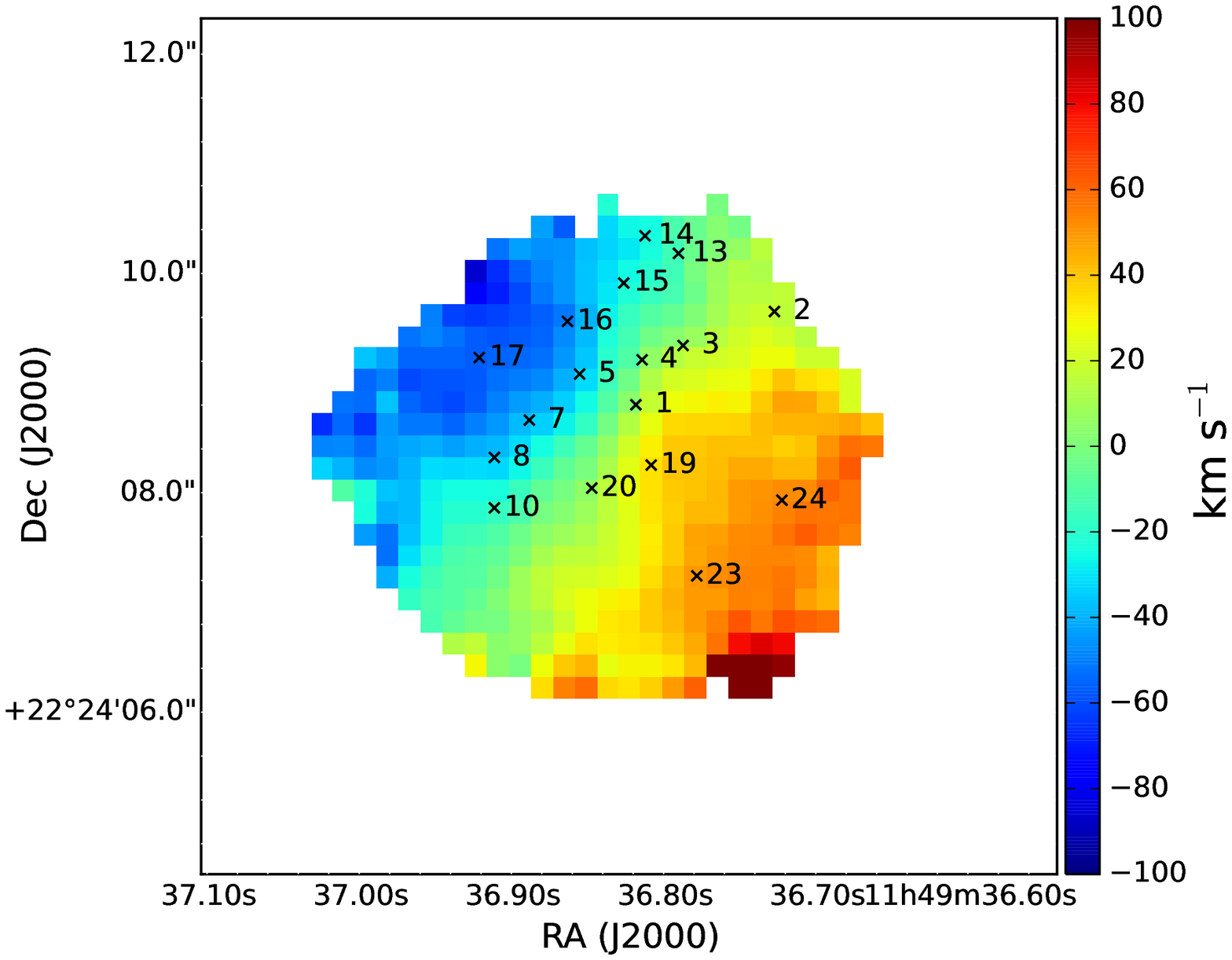}
\includegraphics[width=0.48\textwidth]{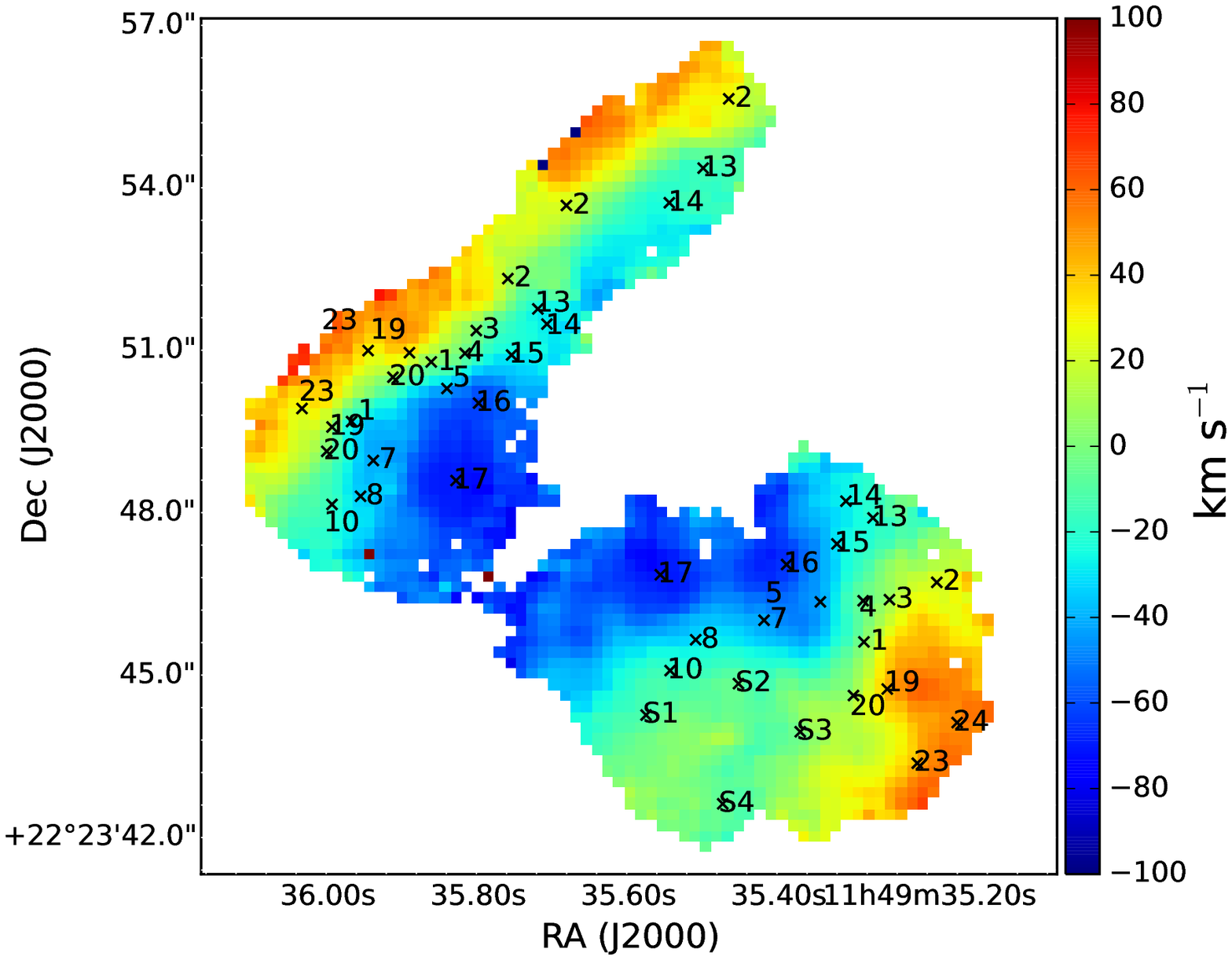}
\caption{Rest frame velocity map of the SN `Refsdal' host derived from the [\ion{O}{2}] line, using a systemic redshift of 1.4888 for the center of the galaxy (knot 1, corresponding to SC in Figures 1 and 2 of \citealt{kar15b}). The velocity is measured after smoothing the datacube by a box of 3.75\arcsec\ on each side. The black crosses correspond to the positions of the knots listed in Table \ref{knots}.}
\label{fi10}
\end{figure}

\begin{figure}
\centering
\includegraphics[width=0.4\textwidth]{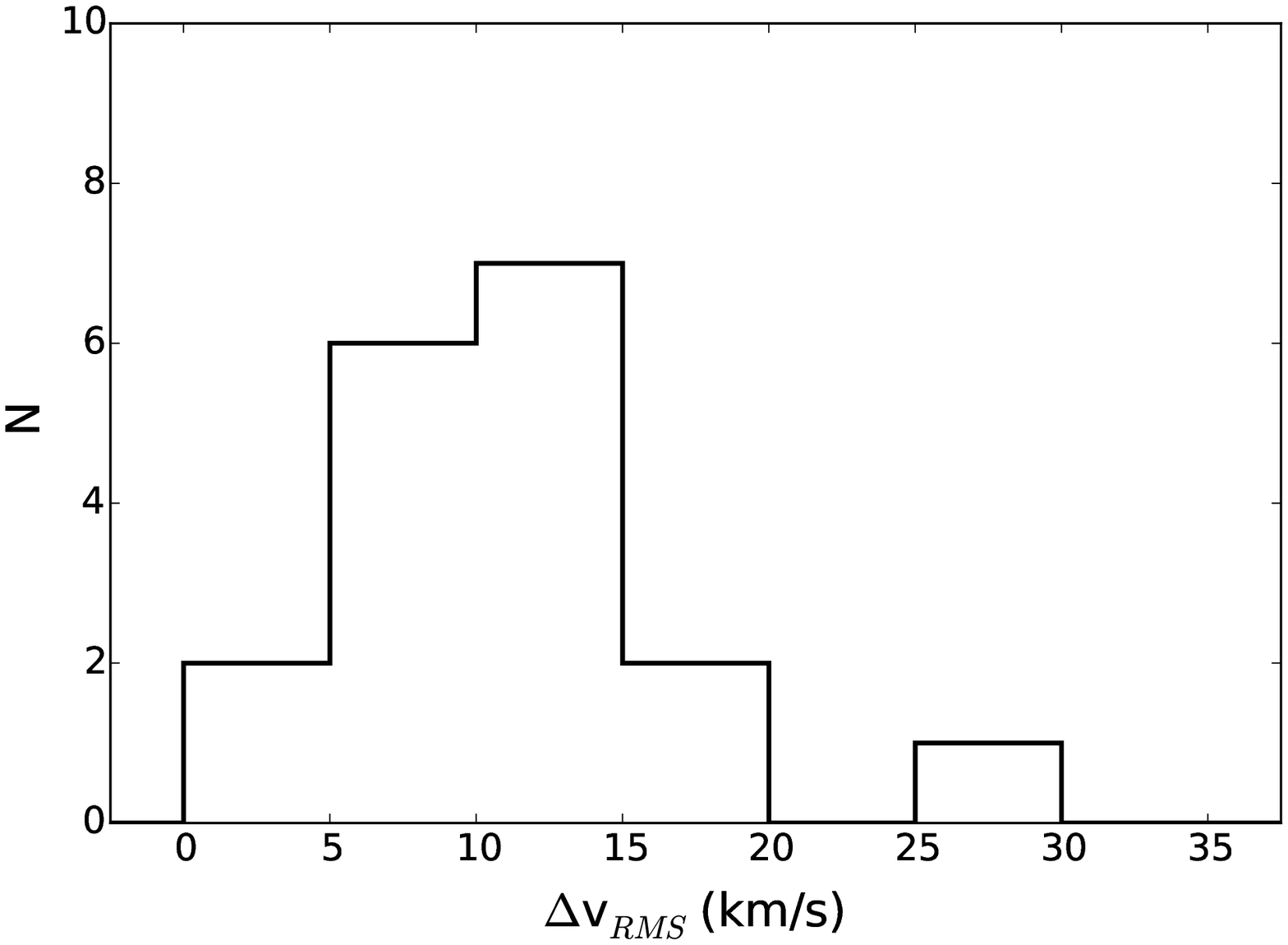}
\caption{Distribution of the standard deviations of the rest frame velocities computed for each of the 18 families of knots in Table \ref{knots}.}
\label{fi11}
\end{figure}

With the large number of multiple image identifications and spectroscopic redshift measurements, we have been able to build a mass model for MACS 1149 that reproduces well the image positions and time delays of the observed images of Refsdal.  Ways to further enhance the strong lens model in the future are to (1) incorporate the stellar velocity dispersion of the $\sim$15 most luminous cluster members that could be measured in the MUSE FoV, (2) extend our present analysis with point-like images to the full surface brightness reconstruction of the Refsdal host and (3) build a dynamical model of the Refsdal host that results consistent with the kinematic observables extracted from the MUSE datacube in spatial pixels corresponding to the same regions on the source plane. In Figure~\ref{fi02}, we show the first comparison between the observed and model-predicted surface brightness of the multiple images of the spiral galaxy in which Refsdal exploded. We note that the predicted surface brightness has been obtained from our current best-fitting model MLV G12F, and does not involve yet any optimization using the extended surface brightness of the multiple images of the spiral galaxy.  We remark that the exquisite \HST\ images taken within the HFF program (the interested reader can find further information on the HFF data release webpages\footnote{http://www.stsci.edu/hst/campaigns/frontier-fields/}) and the power of {\sc Glee} to go beyond the multiple image point-like approximation has enabled our robust identifications of the corresponding Refsdal host knots listed in Table \ref{knots}, which have also been used as inputs in \citet{tre15b}. Moreover, we used the MUSE datacube to produce a rest frame velocity map of the multiple images of the Refsdal host. In Figure~\ref{fi10}, we show the unambiguosly lensed rotation pattern, obtained from the [\ion{O}{2}] emission, of the spiral galaxy at the noteworthy redshift of 1.489. To test the accuracy of the measured velocities, we have performed a bootstrap analysis. By using the variance provided by the MUSE pipeline, we have constructed 100 model datacubes and found that for more than 95\% of the points on the map the measured velocity difference is less than 10 km s$^{-1}$. Starting from this map, we have checked that the velocity values of the corresponding knots of the Refsdal host were consistent. This figure clearly demonstrates the potential of MUSE for integral field spectroscopy and, in particular, for strong lensing analyses. In Figure \ref{fi11}, we show the rms values of the velocities calculated for each of the families of knots listed in Table \ref{knots}. The measured small values further reinforce our selection of corresponding knots.

We defer to future work the next-generation mass reconstruction that incorporates the cluster member velocity dispersions and the surface brightness and velocity field information of the Refsdal host.  This would allow us to investigate in detail the cluster dark-matter halo and cluster member mass distributions (especially on G1 and G2), thus extending our pilot study of MACS J0416.1$-$2403 \citep{gri15} to shed light on the mass substructure properties and the baryonic budget in clusters through comparisons with numerical simulations.

\section{Summary and conclusions}
\label{sec:conclude}

We have presented high-quality VLT/MUSE observations of the central regions of the HFF galaxy cluster MACS 1149, obtained with the main aim of performing a detailed total mass reconstruction of the cluster 
lens. This was a necessary step towards a timely and robust prediction of the next occurrence 
of the SN `Refsdal', the first one to be multiply imaged and resolved at different positions.

The main results of this work can be
summarized as follows:
  
\begin{itemize}

\item[$-$] Within a projected radius of approximately 200 kpc from MACS 1149 center and with a total integration time of 4.8 hours, we have measured secure redshifts for 117 objects. Considering the estimated redshift value of 0.542 for the galaxy cluster, we have identified a large number (68) of cluster members and 18 multiple images of 7 different lensed sources, located at redshifts between 1.240 and 3.703.

\item[$-$] We have reconstructed the measured positions of 88 reliable multiple images from 7 spectroscopic and 3 photometric lensed sources and 18 knots in the SN `Refsdal' host with a remarkably good accuracy, 
i.e.~with 
a rms difference between the observed and model-predicted multiple image positions of only 0.26\arcsec. 
Our best-fitting strong lensing model of MACS 1149 has its total mass distribution parametrized 
in terms of 3 cored elliptical pseudo-isothermal mass components (the most massive one at a projected distance of less than 5 kpc from the BCG luminosity center) and 300 dual elliptical pseudo-isothermal mass components. These represent, respectively, the extended cluster dark-matter halos and the galaxy cluster members, 298 of which have total mass-to-light ratios increasing with the F160W galaxy luminosities and 2 have all model parameters free to vary.

\item[$-$] Including statistical and systematic uncertainties (related only to the different lensing models considered in this study), we predict that the brightness peak of the next appearance of SN `Refsdal' will occur between March and June 2016 and will be approximately 80\% as luminous as the least luminous 
(S4) 
of the already 
detected images, 
thus visible in the coming \HST\ images.

\end{itemize}

We emphasize that the vast amount of information contained in the MUSE and \HST\ data, which we have just started to exploit, will certainly refine the characterizations of the mass properties of MACS 1149. The inclusion of the cluster member internal velocity dispersions from the MUSE spectra and the reconstruction of the full surface brightness distribution in the deep HFF images and velocity map from the MUSE datacube of the SN `Refsdal' host will enable an even more in-depth dissection 
 of the mass composition of this effective lens.

Around the time when our referee report was received, the reappearance of SN `Refsdal' was announced by \citet{kel16}. We note that our original model predictions, reported in \citet{tre15b} and detailed in this article, match very well the observations of the expected image SX, with respect to the position, magnification ratio and time delay derived from the first data of the reappearance.

\acknowledgments
We thank the anonymous refree for useful comments that have helped improve the clarity of the manuscript. C.G. acknowledges support by VILLUM FONDEN Young Investigator Programme through grant no. 10123.
The CLASH Multi-Cycle Treasury Program is based on
observations made with the NASA/ESA {\it Hubble Space Telescope}. The
Space Telescope Science Institute is operated by the Association of
Universities for Research in Astronomy, Inc., under NASA contract NAS
5-26555. ACS was developed under NASA Contract NAS 5-32864. P.R. acknowledges the hospitality and support of the visitor program of the DFG cluster of excellence ``Origin and Structure of the Universe''. This research is supported in part by NASA Grant HST-GO-12065.01-A and HST-GO-13459. 
G.B.C. is supported by the CAPES-ICRANET program through the grant BEX
13946/13-7. We acknowledge support from PRIN-INAF 2014 1.05.01.94.02 (PI: M.~Nonino). 

\clearpage




\clearpage

\appendix
\label{App:AppendixA}
\begin{longtable}{cccccc}
\caption{Catalog of the 300 candidate cluster members.} \\
\hline \hline \noalign{\smallskip}
ID & R.A. & Decl. & F160W & MUSE/WFC3-IR-GRISM/- & Notes \\
 & (J2000) & (J2000) & (mag) & & \\
\noalign{\smallskip} \hline \noalign{\smallskip}
   1 & 11:49:29.546 & +22:23:34.80 & 21.17$\pm$0.01  & -      & \\ 
   2 & 11:49:29.994 & +22:24:16.26 & 23.26$\pm$0.03  & -      & \\ 
   3 & 11:49:30.283 & +22:23:29.97 & 22.24$\pm$0.01  & -      & \\ 
   4 & 11:49:30.337 & +22:24:01.61 & 22.26$\pm$0.02  & -      & \\ 
   5 & 11:49:30.372 & +22:24:18.81 & 20.45$\pm$0.01  & G      & \\ 
   6 & 11:49:30.392 & +22:24:07.47 & 23.30$\pm$0.04  & -      & \\ 
   7 & 11:49:30.460 & +22:23:40.64 & 21.20$\pm$0.01  & -      & \\ 
   8 & 11:49:30.634 & +22:24:11.26 & 21.60$\pm$0.01  & G      & \\ 
   9 & 11:49:30.661 & +22:23:38.45 & 23.75$\pm$0.03  & -      & \\ 
  10 & 11:49:30.891 & +22:23:34.11 & 22.77$\pm$0.02  & -      & \\ 
  11 & 11:49:31.041 & +22:24:32.08 & 22.88$\pm$0.02  & G      & \\ 
  12 & 11:49:31.068 & +22:23:02.00 & 23.38$\pm$0.02  & -      & \\ 
  13 & 11:49:31.117 & +22:23:58.49 & 23.50$\pm$0.02  & -      & \\ 
  14 & 11:49:31.132 & +22:23:42.35 & 22.08$\pm$0.01  & -      & \\ 
  15 & 11:49:31.156 & +22:24:25.41 & 23.05$\pm$0.04  & -      & \\ 
  16 & 11:49:31.165 & +22:24:30.35 & 23.58$\pm$0.05  & -      & \\ 
  17 & 11:49:31.226 & +22:23:00.10 & 23.81$\pm$0.04  & -      & \\ 
  18 & 11:49:31.328 & +22:23:52.42 & 21.30$\pm$0.01  & G      & \\ 
  19 & 11:49:31.462 & +22:24:04.09 & 22.45$\pm$0.01  & G      & \\ 
  20 & 11:49:31.606 & +22:23:49.15 & 19.00$\pm$0.01  & G      & \\ 
  21 & 11:49:31.741 & +22:23:48.66 & 23.48$\pm$0.01  & -      & \\ 
  22 & 11:49:31.788 & +22:23:55.19 & 21.19$\pm$0.01  & -      & \\ 
  23 & 11:49:31.810 & +22:24:25.80 & 22.91$\pm$0.02  & -      & \\ 
  24 & 11:49:31.860 & +22:24:47.15 & 21.13$\pm$0.01  & G      & \\ 
  25 & 11:49:31.861 & +22:23:24.85 & 19.55$\pm$0.01  & G      & \\ 
  26 & 11:49:31.912 & +22:24:24.97 & 21.98$\pm$0.01  & G      & \\ 
  27 & 11:49:31.958 & +22:24:57.31 & 22.52$\pm$0.02  & G      & \\ 
  28 & 11:49:32.012 & +22:24:01.80 & 23.14$\pm$0.03  & -      & \\ 
  29 & 11:49:32.048 & +22:24:53.78 & 21.12$\pm$0.01  & G      & \\ 
  30 & 11:49:32.062 & +22:24:38.60 & 22.74$\pm$0.01  & -      & \\ 
  31 & 11:49:32.093 & +22:25:01.55 & 22.28$\pm$0.01  & -      & \\ 
  32 & 11:49:32.136 & +22:23:58.94 & 22.57$\pm$0.02  & -      & \\ 
  33 & 11:49:32.137 & +22:24:37.03 & 19.70$\pm$0.01  & G      & \\ 
  34 & 11:49:32.157 & +22:23:05.06 & 20.40$\pm$0.01  & G      & \\ 
  35 & 11:49:32.161 & +22:24:24.04 & 21.76$\pm$0.01  & G      & \\ 
  36 & 11:49:32.180 & +22:23:38.99 & 21.27$\pm$0.01  & -      & \\ 
  37 & 11:49:32.190 & +22:23:37.86 & 19.11$\pm$0.01  & G      & \\ 
  38 & 11:49:32.246 & +22:24:33.48 & 23.31$\pm$0.03  & -      & \\ 
  39 & 11:49:32.320 & +22:24:53.86 & 22.88$\pm$0.03  & -      & \\ 
  40 & 11:49:32.320 & +22:23:13.30 & 19.49$\pm$0.01  & G      & \\ 
  41 & 11:49:32.326 & +22:24:10.44 & 23.27$\pm$0.02  & -      & \\ 
  42 & 11:49:32.334 & +22:24:01.77 & 20.03$\pm$0.01  & G      & \\ 
  43 & 11:49:32.342 & +22:23:59.29 & 23.88$\pm$0.04  & -      & \\ 
  44 & 11:49:32.367 & +22:23:55.84 & 21.73$\pm$0.01  & G      & \\ 
  45 & 11:49:32.425 & +22:23:48.15 & 23.88$\pm$0.03  & -      & \\ 
  46 & 11:49:32.716 & +22:25:01.15 & 23.76$\pm$0.04  & -      & \\ 
  47 & 11:49:32.719 & +22:24:12.76 & 23.22$\pm$0.02  & -      & \\ 
  48 & 11:49:32.731 & +22:24:49.35 & 21.93$\pm$0.01  & G      & \\ 
  49 & 11:49:32.789 & +22:24:49.14 & 21.27$\pm$0.01  & G      & \\ 
  50 & 11:49:32.811 & +22:22:37.70 & 23.69$\pm$0.07  & -      & \\ 
  51 & 11:49:32.896 & +22:24:39.74 & 22.12$\pm$0.01  & -      & \\ 
  52 & 11:49:32.964 & +22:23:56.27 & 20.34$\pm$0.01  & G      & \\ 
  53 & 11:49:32.982 & +22:23:40.95 & 19.78$\pm$0.01  & G      & \\ 
  54 & 11:49:33.022 & +22:23:15.29 & 20.97$\pm$0.01  & -      & \\ 
  55 & 11:49:33.022 & +22:23:13.25 & 20.08$\pm$0.01  & G      & \\ 
  56 & 11:49:33.038 & +22:22:51.19 & 23.98$\pm$0.08  & -      & \\ 
  57 & 11:49:33.049 & +22:23:37.50 & 20.91$\pm$0.01  & G      & \\ 
  58 & 11:49:33.075 & +22:23:43.42 & 20.94$\pm$0.01  & G      & \\ 
  59 & 11:49:33.129 & +22:24:42.53 & 20.85$\pm$0.01  & G      & \\ 
  60 & 11:49:33.147 & +22:24:49.54 & 20.02$\pm$0.01  & G      & \\ 
  61 & 11:49:33.148 & +22:24:30.25 & 19.28$\pm$0.01  & G      & \\  
  62 & 11:49:33.189 & +22:24:08.13 & 23.27$\pm$0.02 & -      & \\ 
  63 & 11:49:33.196 & +22:24:42.12 & 21.52$\pm$0.01 & G      & \\ 
  64 & 11:49:33.219 & +22:23:35.74 & 23.96$\pm$0.02 & -      & \\ 
  65 & 11:49:33.223 & +22:23:59.76 & 23.22$\pm$0.02 & -      & \\ 
  66 & 11:49:33.295 & +22:24:01.20 & 23.41$\pm$0.02 & -      & \\ 
  67 & 11:49:33.328 & +22:24:50.94 & 21.12$\pm$0.01 & G      & \\ 
  68 & 11:49:33.360 & +22:23:05.28 & 22.28$\pm$0.01 & G      & \\ 
  69 & 11:49:33.403 & +22:22:55.48 & 23.01$\pm$0.02 & G      & \\ 
  70 & 11:49:33.467 & +22:23:33.40 & 20.64$\pm$0.01 & -      & \\ 
  71 & 11:49:33.468 & +22:23:38.09 & 18.88$\pm$0.01 & M + G      & \\ 
  72 & 11:49:33.473 & +22:24:22.95 & 20.18$\pm$0.01 & G      & \\ 
  73 & 11:49:33.507 & +22:25:00.19 & 23.45$\pm$0.04 & -      & \\ 
  74 & 11:49:33.527 & +22:23:33.74 & 19.48$\pm$0.01 & M + G     & \\ 
  75 & 11:49:33.627 & +22:24:56.55 & 22.35$\pm$0.01 & G      & \\ 
  76 & 11:49:33.637 & +22:24:14.00 & 20.87$\pm$0.01 & M + G     & \\ 
  77 & 11:49:33.671 & +22:24:50.32 & 23.27$\pm$0.02 & -      & \\ 
  78 & 11:49:33.682 & +22:22:50.77 & 20.51$\pm$0.01 & G      & \\ 
  79 & 11:49:33.721 & +22:23:07.32 & 22.49$\pm$0.02 & -      & \\ 
  80 & 11:49:33.776 & +22:24:33.97 & 23.60$\pm$0.02 & -      & \\ 
  81 & 11:49:33.832 & +22:24:06.05 & 20.11$\pm$0.01 & M + G     & \\ 
  82 & 11:49:33.851 & +22:24:11.82 & 23.16$\pm$0.01 & -      & \\ 
  83 & 11:49:33.863 & +22:24:17.68 & 19.53$\pm$0.01 & M + G     & \\ 
  84 & 11:49:33.889 & +22:23:33.77 & 19.94$\pm$0.01 & M + G     & \\ 
  85 & 11:49:33.921 & +22:24:28.05 & 22.08$\pm$0.01 & G      & \\ 
  86 & 11:49:33.932 & +22:24:03.85 & 21.57$\pm$0.01 & M + G     & \\ 
  87 & 11:49:33.946 & +22:22:46.26 & 22.48$\pm$0.02 & -      & \\ 
  88 & 11:49:34.003 & +22:23:26.22 & 19.78$\pm$0.01 & M + G     & \\ 
  89 & 11:49:34.038 & +22:24:19.04 & 20.18$\pm$0.01 & M + G     & \\ 
  90 & 11:49:34.105 & +22:24:12.28 & 23.28$\pm$0.02 & -      & \\ 
  91 & 11:49:34.120 & +22:24:04.61 & 21.93$\pm$0.01 & M      & \\ 
  92 & 11:49:34.125 & +22:24:53.14 & 23.04$\pm$0.02 & -      & \\ 
  93 & 11:49:34.214 & +22:24:38.13 & 21.38$\pm$0.01 & -      & \\ 
  94 & 11:49:34.240 & +22:23:33.83 & 19.29$\pm$0.01 & M + G     & \\ 
  95 & 11:49:34.245 & +22:23:39.70 & 20.74$\pm$0.01 & M + G     & \\ 
  96 & 11:49:34.267 & +22:23:53.14 & 19.60$\pm$0.01 & M + G     & \\ 
  97 & 11:49:34.291 & +22:25:05.51 & 21.85$\pm$0.01 & G      & \\ 
  98 & 11:49:34.291 & +22:23:49.57 & 20.25$\pm$0.01 & M + G     & \\ 
  99 & 11:49:34.298 & +22:23:35.09 & 21.93$\pm$0.01 & -      & \\ 
 100 & 11:49:34.305 & +22:24:42.74 & 19.14$\pm$0.01 & G      & \\ 
 101 & 11:49:34.353 & +22:24:40.82 & 20.05$\pm$0.01 & G      & \\ 
 102 & 11:49:34.401 & +22:24:37.55 & 22.40$\pm$0.01 & G      & \\ 
 103 & 11:49:34.432 & +22:22:47.33 & 23.99$\pm$0.04 & -      & \\ 
 104 & 11:49:34.439 & +22:25:13.47 & 23.73$\pm$0.03 & -      & \\ 
 105 & 11:49:34.475 & +22:23:41.36 & 22.89$\pm$0.01 & -      & \\ 
 106 & 11:49:34.515 & +22:24:08.30 & 21.98$\pm$0.01 & M + G     & \\ 
 107 & 11:49:34.518 & +22:24:42.10 & 19.09$\pm$0.01 & G      & \\ 
 108 & 11:49:34.563 & +22:23:43.20 & 23.38$\pm$0.02 & -      & \\ 
 109 & 11:49:34.572 & +22:24:30.37 & 20.42$\pm$0.01 & G      & \\ 
 110 & 11:49:34.579 & +22:24:45.80 & 21.63$\pm$0.01 & -      & \\ 
 111 & 11:49:34.598 & +22:23:42.12 & 22.69$\pm$0.01 & M      & \\ 
 112 & 11:49:34.610 & +22:24:44.88 & 20.99$\pm$0.01 & G      & \\ 
 113 & 11:49:34.628 & +22:24:27.18 & 21.36$\pm$0.01 & G      & \\ 
 114 & 11:49:34.631 & +22:23:03.44 & 23.45$\pm$0.01 & -      & \\ 
 115 & 11:49:34.688 & +22:24:02.28 & 21.81$\pm$0.01 & M      & \\ 
 116 & 11:49:34.720 & +22:24:47.92 & 23.41$\pm$0.02 & -      & \\ 
 117 & 11:49:34.725 & +22:22:43.04 & 21.42$\pm$0.01 & -      & \\ 
 118 & 11:49:34.733 & +22:24:40.94 & 22.27$\pm$0.01 & -      & \\ 
 119 & 11:49:34.735 & +22:24:51.88 & 20.48$\pm$0.01 & G      & \\ 
 120 & 11:49:34.761 & +22:23:34.53 & 22.24$\pm$0.01 & M      & \\ 
 121 & 11:49:34.807 & +22:23:45.66 & 21.64$\pm$0.01 & M + G     & \\ 
 122 & 11:49:34.818 & +22:23:23.47 & 22.05$\pm$0.01 & M      & \\ 
 123 & 11:49:34.844 & +22:22:54.67 & 23.05$\pm$0.01 & -      & \\ 
 124 & 11:49:34.857 & +22:24:53.76 & 20.45$\pm$0.01 & G      & \\ 
 125 & 11:49:34.858 & +22:23:49.17 & 23.68$\pm$0.02 & -      & \\ 
 126 & 11:49:34.865 & +22:24:03.78 & 21.21$\pm$0.01 & M + G     & \\ 
 127 & 11:49:34.918 & +22:23:15.62 & 21.36$\pm$0.01 & G      & \\ 
 128 & 11:49:34.992 & +22:25:05.59 & 23.65$\pm$0.02 & G      & \\ 
 129 & 11:49:35.001 & +22:23:36.60 & 20.47$\pm$0.01 & M + G     & \\ 
 130 & 11:49:35.064 & +22:23:02.75 & 22.19$\pm$0.01 & -      & \\ 
 131 & 11:49:35.160 & +22:23:52.53 & 23.27$\pm$0.01 & -      & \\ 
 132 & 11:49:35.182 & +22:24:47.98 & 19.65$\pm$0.01 & G      & \\ 
 133 & 11:49:35.224 & +22:23:01.62 & 22.64$\pm$0.01 & -      & \\ 
 134 & 11:49:35.230 & +22:22:42.59 & 21.65$\pm$0.01 & -      & \\ 
 135 & 11:49:35.235 & +22:24:36.04 & 22.85$\pm$0.02 & -      & \\ 
 126 & 11:49:35.237 & +22:24:38.65 & 23.72$\pm$0.02 & -      & \\ 
 137 & 11:49:35.241 & +22:25:02.87 & 19.33$\pm$0.01 & G      & \\ 
 138 & 11:49:35.251 & +22:23:32.30 & 20.90$\pm$0.01 & M      & \\ 
 139 & 11:49:35.259 & +22:22:39.86 & 22.08$\pm$0.01 & -      & \\ 
 140 & 11:49:35.260 & +22:25:05.17 & 22.49$\pm$0.01 & G      & \\ 
 141 & 11:49:35.265 & +22:23:34.72 & 19.82$\pm$0.01 & M + G      & \\ 
 142 & 11:49:35.288 & +22:24:41.63 & 20.95$\pm$0.01 & G      & \\ 
 143 & 11:49:35.291 & +22:24:27.33 & 24.58$\pm$0.04 & G      & \\ 
 144 & 11:49:35.343 & +22:24:01.05 & 21.71$\pm$0.01 & M      & \\ 
 145 & 11:49:35.369 & +22:23:22.60 & 21.74$\pm$0.01 & -      & \\ 
 146 & 11:49:35.405 & +22:23:58.30 & 19.69$\pm$0.01 & M + G      & \\ 
 147 & 11:49:35.429 & +22:24:10.36 & 22.06$\pm$0.01 & M      & \\ 
 148 & 11:49:35.444 & +22:24:58.27 & 21.51$\pm$0.01 & G      & \\ 
 149 & 11:49:35.453 & +22:24:54.77 & 21.30$\pm$0.01 & G      & \\ 
 150 & 11:49:35.469 & +22:23:43.63 & 19.65$\pm$0.01 & M + G  & G1   \\ 
 151 & 11:49:35.498 & +22:23:49.52 & 23.44$\pm$0.01 & -      & \\ 
 152 & 11:49:35.500 & +22:24:14.18 & 22.35$\pm$0.01 & M + G     & \\ 
 153 & 11:49:35.509 & +22:24:03.77 & 19.32$\pm$0.01 & M + G     & \\ 
 154 & 11:49:35.514 & +22:24:16.48 & 23.12$\pm$0.02 & -      & \\ 
 155 & 11:49:35.553 & +22:24:57.45 & 22.29$\pm$0.01 & -      & \\ 
 156 & 11:49:35.559 & +22:24:26.67 & 19.75$\pm$0.01 & G      & \\ 
 157 & 11:49:35.568 & +22:24:58.61 & 21.40$\pm$0.01 & G      & \\ 
 158 & 11:49:35.584 & +22:24:38.84 & 19.33$\pm$0.01 & G      & \\ 
 159 & 11:49:35.593 & +22:24:11.29 & 23.16$\pm$0.02 & -      & \\ 
 160 & 11:49:35.593 & +22:24:57.68 & 22.55$\pm$0.01 & -      & \\ 
 161 & 11:49:35.629 & +22:24:19.31 & 21.66$\pm$0.01 & M + G     & \\ 
 162 & 11:49:35.652 & +22:23:23.24 & 21.37$\pm$0.01 & M      & \\ 
 163 & 11:49:35.665 & +22:23:53.09 & 21.00$\pm$0.01 & M      & \\ 
 164 & 11:49:35.665 & +22:24:25.81 & 23.01$\pm$0.02 & -      & \\ 
 165 & 11:49:35.686 & +22:23:32.29 & 20.11$\pm$0.01 & M + G     & \\ 
 166 & 11:49:35.699 & +22:23:54.71 & 17.99$\pm$0.01 & M + G & BCG   \\ 
 167 & 11:49:35.727 & +22:24:06.52 & 20.75$\pm$0.01 & M + G     & \\ 
 168 & 11:49:35.761 & +22:25:06.22 & 20.23$\pm$0.01 & G      & \\ 
 169 & 11:49:35.763 & +22:22:48.70 & 20.12$\pm$0.01 & G      & \\ 
 170 & 11:49:35.806 & +22:24:03.25 & 23.16$\pm$0.02 & M      & \\ 
 171 & 11:49:35.829 & +22:22:53.12 & 19.15$\pm$0.01 & -      & \\ 
 172 & 11:49:35.845 & +22:24:47.89 & 22.83$\pm$0.01 & -      & \\ 
 173 & 11:49:35.864 & +22:24:55.60 & 19.44$\pm$0.01 & G      & \\ 
 174 & 11:49:35.881 & +22:24:44.21 & 22.57$\pm$0.01 & -      & \\ 
 175 & 11:49:35.903 & +22:24:08.32 & 23.20$\pm$0.02 & -      & \\ 
 176 & 11:49:35.908 & +22:25:15.28 & 22.96$\pm$0.01 & G      & \\ 
 177 & 11:49:35.917 & +22:23:58.59 & 21.20$\pm$0.01 & M      & \\ 
 178 & 11:49:35.943 & +22:25:06.49 & 22.56$\pm$0.02 & -      & \\ 
 179 & 11:49:35.952 & +22:24:53.67 & 19.19$\pm$0.01 & G      & \\ 
 180 & 11:49:35.958 & +22:23:50.13 & 19.44$\pm$0.01 & M + G & G2 \\ 
 181 & 11:49:36.045 & +22:22:45.24 & 23.96$\pm$0.05 & -      & \\ 
 182 & 11:49:36.048 & +22:23:39.89 & 22.86$\pm$0.01 & M      & \\ 
 183 & 11:49:36.052 & +22:23:52.57 & 21.30$\pm$0.01 & -      & \\ 
 184 & 11:49:36.097 & +22:23:53.54 & 21.68$\pm$0.01 & M      & \\ 
 185 & 11:49:36.188 & +22:23:46.49 & 21.78$\pm$0.01 & M + G     & \\ 
 186 & 11:49:36.192 & +22:23:37.59 & 22.79$\pm$0.01 & M      & \\ 
 187 & 11:49:36.247 & +22:23:52.36 & 19.46$\pm$0.01 & M + G     & \\ 
 188 & 11:49:36.289 & +22:24:01.20 & 19.72$\pm$0.01 & M + G     & \\ 
 189 & 11:49:36.303 & +22:24:56.16 & 23.10$\pm$0.02 & -      & \\ 
 190 & 11:49:36.304 & +22:25:07.92 & 20.78$\pm$0.01 & G      & \\ 
 191 & 11:49:36.333 & +22:24:39.58 & 23.48$\pm$0.03 & -      & \\ 
 192 & 11:49:36.382 & +22:24:40.70 & 21.52$\pm$0.01 & G      & \\ 
 193 & 11:49:36.396 & +22:24:34.10 & 22.55$\pm$0.01 & -      & \\ 
 194 & 11:49:36.414 & +22:23:55.69 & 21.86$\pm$0.01 & M      & \\ 
 195 & 11:49:36.503 & +22:22:40.43 & 23.34$\pm$0.03 & -      & \\ 
 196 & 11:49:36.519 & +22:22:58.19 & 23.79$\pm$0.02 & G      & \\ 
 197 & 11:49:36.541 & +22:23:59.09 & 20.32$\pm$0.01 & M + G     & \\ 
 198 & 11:49:36.546 & +22:24:40.53 & 23.03$\pm$0.03 & -      & \\ 
 199 & 11:49:36.574 & +22:23:52.92 & 22.63$\pm$0.01 & M      & \\ 
 200 & 11:49:36.580 & +22:22:33.72 & 22.79$\pm$0.02 & -      & \\ 
 201 & 11:49:36.607 & +22:24:37.67 & 21.08$\pm$0.01 & G      & \\ 
 202 & 11:49:36.608 & +22:25:07.97 & 21.55$\pm$0.01 & G      & \\ 
 203 & 11:49:36.628 & +22:23:46.26 & 20.36$\pm$0.01 & M + G     & \\ 
 204 & 11:49:36.643 & +22:23:47.17 & 22.89$\pm$0.01 & -      & \\ 
 205 & 11:49:36.644 & +22:22:43.36 & 23.08$\pm$0.03 & -      & \\ 
 206 & 11:49:36.692 & +22:24:07.23 & 22.12$\pm$0.01 & M + G     & \\ 
 207 & 11:49:36.699 & +22:22:40.48 & 21.60$\pm$0.01 & -      & \\ 
 208 & 11:49:36.704 & +22:24:42.06 & 21.80$\pm$0.01 & G      & \\ 
 209 & 11:49:36.736 & +22:24:15.79 & 20.55$\pm$0.01 & M + G    & \\ 
 210 & 11:49:36.756 & +22:24:52.13 & 22.71$\pm$0.01 & -      & \\ 
 211 & 11:49:36.796 & +22:23:49.92 & 23.86$\pm$0.02 & -      & \\ 
 212 & 11:49:36.825 & +22:22:40.39 & 21.52$\pm$0.01 & -      & \\ 
 213 & 11:49:36.858 & +22:23:46.98 & 18.89$\pm$0.01 & M + G     & \\ 
 214 & 11:49:36.860 & +22:24:20.17 & 23.16$\pm$0.01 & -      & \\ 
 215 & 11:49:36.878 & +22:23:30.99 & 20.32$\pm$0.01 & M + G     & \\ 
 216 & 11:49:36.881 & +22:23:38.63 & 23.79$\pm$0.03 & -      & \\ 
 217 & 11:49:36.883 & +22:23:18.22 & 21.15$\pm$0.01 & G      & \\ 
 218 & 11:49:36.886 & +22:23:20.79 & 20.08$\pm$0.01 & M + G     & \\ 
 219 & 11:49:36.892 & +22:24:16.47 & 19.67$\pm$0.01 & M + G     & \\ 
 220 & 11:49:36.924 & +22:23:20.75 & 23.06$\pm$0.01 & -      & \\ 
 221 & 11:49:36.964 & +22:24:07.68 & 22.30$\pm$0.01 & M      & \\ 
 222 & 11:49:36.968 & +22:24:10.88 & 21.50$\pm$0.01 & M      & \\ 
 223 & 11:49:36.972 & +22:23:10.96 & 19.97$\pm$0.01 & G      & \\ 
 224 & 11:49:36.990 & +22:23:07.17 & 21.70$\pm$0.01 & G      & \\ 
 225 & 11:49:37.100 & +22:23:47.16 & 23.06$\pm$0.02 & -      & \\ 
 226 & 11:49:37.135 & +22:22:56.00 & 20.67$\pm$0.01 & G      & \\ 
 227 & 11:49:37.162 & +22:25:00.45 & 22.29$\pm$0.02 & -      & \\ 
 228 & 11:49:37.191 & +22:24:39.12 & 23.31$\pm$0.02 & -      & \\ 
 229 & 11:49:37.231 & +22:23:53.07 & 23.99$\pm$0.02 & -      & \\ 
 230 & 11:49:37.238 & +22:23:59.18 & 21.90$\pm$0.01 & M + G     & \\ 
 231 & 11:49:37.288 & +22:23:29.95 & 22.51$\pm$0.01 & M      & \\ 
 232 & 11:49:37.306 & +22:23:52.32 & 19.66$\pm$0.01 & M + G     & \\ 
 233 & 11:49:37.442 & +22:24:32.01 & 23.03$\pm$0.02 & -      & \\ 
 234 & 11:49:37.445 & +22:22:51.75 & 22.02$\pm$0.01 & -      & \\ 
 235 & 11:49:37.469 & +22:23:56.82 & 22.86$\pm$0.01 & -      & \\ 
 236 & 11:49:37.509 & +22:24:19.45 & 21.74$\pm$0.01 & M + G     & \\ 
 237 & 11:49:37.518 & +22:24:23.65 & 22.10$\pm$0.02 & -      & \\ 
 238 & 11:49:37.549 & +22:23:22.50 & 18.09$\pm$0.01 & M + G     & Ref \\ 
 239 & 11:49:37.588 & +22:24:16.19 & 23.87$\pm$0.02 & -      & \\ 
 240 & 11:49:37.592 & +22:23:43.97 & 21.45$\pm$0.01 & M + G     & \\ 
 241 & 11:49:37.619 & +22:23:14.82 & 20.39$\pm$0.01 & G      & \\ 
 242 & 11:49:37.656 & +22:23:44.95 & 20.68$\pm$0.01 & M + G     & \\ 
 243 & 11:49:37.716 & +22:23:12.08 & 23.35$\pm$0.03 & -      & \\ 
 244 & 11:49:37.744 & +22:23:29.20 & 22.54$\pm$0.01 & M + G     & \\ 
 245 & 11:49:37.745 & +22:23:52.42 & 23.46$\pm$0.02 & -      & \\ 
 246 & 11:49:37.770 & +22:23:41.25 & 23.76$\pm$0.03 & M + G     & \\ 
 247 & 11:49:37.793 & +22:23:56.83 & 22.65$\pm$0.01 & M + G     & \\ 
 248 & 11:49:37.804 & +22:24:10.99 & 18.76$\pm$0.01 & M + G     & \\ 
 249 & 11:49:38.003 & +22:24:27.79 & 20.34$\pm$0.01 & G      & \\ 
 250 & 11:49:38.039 & +22:23:35.57 & 21.53$\pm$0.01 & G      & \\ 
 251 & 11:49:38.129 & +22:23:47.91 & 23.82$\pm$0.02 & -      & \\ 
 252 & 11:49:38.181 & +22:24:55.80 & 20.89$\pm$0.01 & G      & \\ 
 253 & 11:49:38.211 & +22:23:39.75 & 22.75$\pm$0.02 & -      & \\ 
 254 & 11:49:38.269 & +22:23:43.30 & 22.23$\pm$0.01 & -      & \\ 
 255 & 11:49:38.282 & +22:23:34.23 & 21.02$\pm$0.01 & G      & \\ 
 256 & 11:49:38.372 & +22:23:52.21 & 22.45$\pm$0.01 & -      & \\ 
 257 & 11:49:38.397 & +22:24:48.64 & 20.29$\pm$0.01 & G      & \\ 
 258 & 11:49:38.430 & +22:23:21.84 & 19.86$\pm$0.01 & G      & \\ 
 259 & 11:49:38.430 & +22:24:00.69 & 23.58$\pm$0.03 & -      & \\ 
 260 & 11:49:38.502 & +22:23:02.54 & 19.83$\pm$0.01 & G      & \\ 
 261 & 11:49:38.503 & +22:23:09.34 & 22.69$\pm$0.01 & G      & \\ 
 262 & 11:49:38.513 & +22:22:48.86 & 21.59$\pm$0.01 & -      & \\ 
 263 & 11:49:38.526 & +22:24:28.26 & 23.48$\pm$0.02 & -      & \\ 
 264 & 11:49:38.553 & +22:23:42.45 & 21.40$\pm$0.01 & -      & \\ 
 265 & 11:49:38.570 & +22:23:09.22 & 19.49$\pm$0.01 & G      & \\ 
 266 & 11:49:38.829 & +22:23:25.80 & 22.28$\pm$0.01 & -      & \\ 
 267 & 11:49:38.836 & +22:23:45.77 & 21.64$\pm$0.01 & G      & \\ 
 268 & 11:49:38.849 & +22:22:56.72 & 19.37$\pm$0.01 & G      & \\ 
 269 & 11:49:38.882 & +22:23:51.38 & 21.15$\pm$0.01 & -      & \\ 
 270 & 11:49:38.928 & +22:22:59.90 & 18.51$\pm$0.01 & G      & \\ 
 271 & 11:49:39.004 & +22:22:57.05 & 22.48$\pm$0.02 & -      & \\ 
 272 & 11:49:39.091 & +22:23:28.38 & 21.77$\pm$0.01 & -      & \\ 
 273 & 11:49:39.092 & +22:23:37.15 & 21.33$\pm$0.01 & G      & \\ 
 274 & 11:49:39.165 & +22:24:58.28 & 23.81$\pm$0.03 & -      & \\ 
 275 & 11:49:39.188 & +22:23:59.85 & 22.90$\pm$0.02 & -      & \\ 
 276 & 11:49:39.228 & +22:24:46.00 & 23.65$\pm$0.03 & -      & \\ 
 277 & 11:49:39.310 & +22:23:52.48 & 21.06$\pm$0.01 & G      & \\ 
 278 & 11:49:39.320 & +22:24:30.43 & 18.78$\pm$0.01 & G      & \\ 
 279 & 11:49:39.376 & +22:23:01.02 & 20.18$\pm$0.01 & G      & \\ 
 280 & 11:49:39.418 & +22:24:50.46 & 23.06$\pm$0.03 & -      & \\ 
 281 & 11:49:39.431 & +22:23:24.98 & 22.42$\pm$0.02 & -      & \\ 
 282 & 11:49:39.514 & +22:23:39.56 & 20.04$\pm$0.01 & G      & \\ 
 283 & 11:49:39.565 & +22:23:05.35 & 22.98$\pm$0.02 & -      & \\ 
 284 & 11:49:39.656 & +22:24:08.30 & 23.64$\pm$0.03 & -      & \\ 
 285 & 11:49:39.824 & +22:23:49.01 & 22.90$\pm$0.01 & -      & \\ 
 286 & 11:49:39.899 & +22:23:32.39 & 20.39$\pm$0.01 & G      & \\ 
 287 & 11:49:39.908 & +22:23:22.25 & 22.47$\pm$0.04 & -      & \\ 
 288 & 11:49:40.163 & +22:24:04.04 & 19.23$\pm$0.01 & G      & \\ 
 289 & 11:49:40.195 & +22:23:59.29 & 18.98$\pm$0.01 & G      & \\ 
 290 & 11:49:40.218 & +22:23:26.55 & 23.60$\pm$0.06 & -      & \\ 
 291 & 11:49:40.273 & +22:24:54.65 & 19.94$\pm$0.01 & -      & \\ 
 292 & 11:49:40.360 & +22:24:26.56 & 22.36$\pm$0.01 & -      & \\ 
 293 & 11:49:40.366 & +22:23:57.60 & 20.34$\pm$0.01 & -      & \\ 
 294 & 11:49:40.473 & +22:23:22.52 & 19.20$\pm$0.01 & G      & \\ 
 295 & 11:49:40.485 & +22:24:56.62 & 21.49$\pm$0.01 & -      & \\ 
 296 & 11:49:40.573 & +22:24:19.34 & 23.78$\pm$0.03 & -      & \\ 
 297 & 11:49:40.778 & +22:24:24.15 & 22.96$\pm$0.02 & -      & \\ 
 298 & 11:49:40.833 & +22:23:39.92 & 21.80$\pm$0.01 & G      & \\ 
 299 & 11:49:40.987 & +22:23:24.34 & 19.21$\pm$0.01 & -      & \\ 
 300 & 11:49:41.368 & +22:24:11.84 & 23.42$\pm$0.02 & -       & \\ 
\noalign{\smallskip} \hline
\label{tab10}
\end{longtable}

\clearpage

\end{document}